\documentstyle[twoside,fleqn,epsfig]{article}

\def\be{\begin{equation}}
\def\ee{\end{equation}}
\def\bea{\begin{eqnarray}}
\def\eea{\end{eqnarray}}
\newcommand{\lsim}{\mathrel{\mathop{\kern 0pt \rlap
  {\raise.2ex\hbox{$<$}}} \lower.9ex\hbox{\kern-.190em $\sim$}}}
\newcommand{\gsim}{\mathrel{\mathop{\kern 0pt \rlap
  {\raise.2ex\hbox{$>$}}}
  \lower.9ex\hbox{\kern-.190em $\sim$}}}

\newcommand{\AmS}{{\protect\the\textfont2
  A\kern-.1667em\lower.5ex\hbox{M}\kern-.125emS}}
\normalsize
\begin{document}

\baselineskip=0.65cm

\begin{center}
\Large
{\bf First model independent results from DAMA/LIBRA--phase2}
\rm
\end{center}

\large
\vspace{0.4cm}

\begin{center}

R.\,Bernabei$^{a,b}$,~P.\,Belli$^{a,b}$,~A.\,Bussolotti$^{b}$,~F.\,Cappella$^{c,d}$,
\vspace{1mm}

V.\,Caracciolo$^{e}$,~R.\,Cerulli$^{a,b}$,~C.J.\,Dai$^{f}$,~A.\,d'Angelo$^{c,d}$,
\vspace{1mm}

~A. Di Marco$^{b}$,~H.L.\,He$^{f}$, A.\,Incicchitti$^{c,d}$,
\vspace{1mm}

~X.H.\,Ma$^{f}$,~A.\,Mattei$^{d}$,~V.\,Merlo$^{a,b}$,~F.\,Montecchia$^{b,g}$,
\vspace{1mm}

~X.D.\,Sheng$^{f}$,~Z.P.\,Ye$^{f,h}$
\vspace{1mm}

\normalsize
\vspace{0.4cm}

$^{a}${\it Dip. di Fisica, Universit\`a di Roma ``Tor Vergata'', Rome, Italy}
\vspace{1mm}

$^{b}${\it INFN, sez. Roma ``Tor Vergata'', Rome, Italy}
\vspace{1mm}

$^{c}${\it Dip. di Fisica, Universit\`a di Roma ``La Sapienza'', Rome, Italy}
\vspace{1mm}

$^{d}${\it INFN, sez. Roma, Rome, Italy}
\vspace{1mm}

$^{e}${\it INFN Laboratori Nazionali del Gran Sasso, Assergi, Italy}
\vspace{1mm}

$^{f}${\it Key Laboratory of Particle Astrophysics, Institute of High Energy Physics, \\
Chinese Academy of Sciences, Beijing, China}
\vspace{1mm}

$^{g}${\it Dip. Ingegneria Civile e Ingegneria Informatica, Universit\`a di Roma \\
``Tor Vergata'', Rome, Italy}
\vspace{1mm}

$^{h}${\it University of Jinggangshan, Ji'an, Jiangxi, P.R. China}

\end{center}
	
\normalsize

\begin{abstract}

The first model independent results obtained by the DAMA/LIBRA--phase2 experiment
are presented. The data have been collected over 6 annual cycles 
corresponding to a total exposure of 1.13 ton $\times$ yr, deep underground at the 
Gran Sasso National Laboratory (LNGS) of the I.N.F.N.
The DAMA/LIBRA--phase2 apparatus, $\simeq$ 250 kg highly radio-pure NaI(Tl), profits
from a second generation high quantum efficiency photomultipliers and of new electronics
with respect to DAMA/LIBRA--phase1. The improved experimental configuration
has also allowed to lower the software energy threshold.
New data analysis strategies are presented.
The DAMA/LIBRA--phase2 data confirm the evidence of a signal that meets all the 
requirements of the model independent Dark Matter (DM) annual modulation signature,
at 9.5 $\sigma$ C.L. in the energy region (1--6) keV. 
In the energy region between 2 and 6 keV, where data are
also available from DAMA/NaI and DAMA/LIBRA--phase1
(exposure $1.33$ ton $\times$ yr, collected over 14 annual cycles),
the achieved C.L. for the full exposure (2.46 ton $\times$ yr) is 12.9 $\sigma$;
the modulation amplitude of the {\it single-hit} scintillation events 
is: $(0.0103 \pm 0.0008)$ cpd/kg/keV,
the measured phase is $(145 \pm 5)$ days 
and the measured period is $(0.999 \pm 0.001)$ yr, 
all these values are well in agreement with those expected for DM particles. 
No systematics or side reaction able to mimic the exploited DM signature 
(i.e. to account for the whole measured modulation amplitude and  to 
simultaneously satisfy all the requirements of the signature), has 
been found or suggested by anyone throughout some decades thus far.
\end{abstract}

\vspace{5.0mm}


\vspace{2.0mm}


\section{Introduction}

The DAMA/LIBRA 
\cite{perflibra,modlibra,modlibra2,modlibra3,review,pmts,mu,norole,daissue,diurna,papep,cnc-l,IPP,shadow,bot11,mirasim,model,mirsim,uni18,bled18}
experiment,
as the pioneer DAMA/NaI 
\cite{prop,allDM1,allDM2,allDM3,allDM4,allDM5,allDM6,allDM7,allDM8,Nim98,Sist,RNC,ijmd,ijma,epj06,ijma07,chan,wimpele,ldm,
allRare1,allRare2,allRare3,allRare4,allRare5,allRare6,allRare7,allRare8,allRare9},
has the main aim to investigate the presence of DM particles in the galactic halo by exploiting 
the DM annual modulation signature (originally suggested in Ref.~\cite{Freese1,Freese2}). 
In addition, the developed highly radio-pure NaI(Tl) target-detectors 
\cite{perflibra,pmts,daissue,ULBNaI} ensure sensitivity to a wide range 
of DM candidates, interaction types and astrophysical scenarios 
(see e.g. Refs. 
\cite{modlibra,shadow,mirasim,model,mirsim,allDM1,allDM2,allDM3,allDM4,allDM5,allDM6,allDM7,allDM8,RNC,ijmd,ijma,epj06,ijma07,chan,wimpele,ldm}, and 
in literature).

\vspace{0.1cm}

The origin of the DM annual modulation signature and of its peculiar features  is
due to the Earth motion with respect to the DM particles constituting the Galactic Dark Halo,
so it is not related to terrestrial seasons.
In fact, as a consequence of the Earth's revolution around the Sun, 
which is moving in the Galaxy with respect to the Local Standard of 
Rest towards the star Vega near
the constellation of Hercules, the Earth should be crossed
by a larger flux of DM particles around $\simeq$ 2 June
and by a smaller one around $\simeq$ 2 December.
In the former case the Earth orbital velocity is summed to that of the
solar system with respect to the Galaxy, while in the latter
the two velocities are subtracted.
The DM annual modulation signature is very distinctive since the effect
induced by DM particles must simultaneously satisfy
all the following requirements: the rate must contain a component
modulated according to a cosine function (1) with
one year period (2) and a phase that peaks roughly 
$\simeq$ 2 June (3); this modulation must only be found in a
well-defined low energy range, where DM particle induced
events can be present (4); it must apply only to those events
in which just one detector of many actually ``fires'' ({\it single-hit}
 events), since the DM particle multi-interaction probability
is negligible (5); the modulation amplitude in the region
of maximal sensitivity must be $\lsim$ 7\% 
of the constant part of the signal 
for usually 
adopted
halo distributions (6), but it can be larger in case of some
proposed scenarios such as e.g. those in Ref.~\cite{Wei01_1,Wei01_2,Wei01_3,Fre04_1,Fre04_2} 
(even up to $\simeq$ 30\%).
Thus this signature is not dependent on the nature of the DM particle, 
has many peculiarities and, in addition, it allows to test a wide range 
of parameters in many possible astrophysical, nuclear and particle 
physics scenarios.

This DM signature might be mimicked only by systematic effects or side reactions 
able to account for the whole observed modulation amplitude and
to simultaneously satisfy all the requirements given above;
none able to do that has been found or suggested by anyone throughout
some decades thus far (see e.g. Ref. \cite{perflibra,modlibra,modlibra2,modlibra3,review,mu,norole,Sist,RNC,ijmd,uni18}.

The full description of the DAMA/LIBRA set-up and the adopted procedures 
during the phase1 and other related arguments have been discussed in 
details e.g. in Refs. \cite{perflibra,modlibra,modlibra2,modlibra3,review}.

At the end of 2010 the upgrade of DAMA/LIBRA--phase2 started. All the photomultipliers (PMTs) 
were replaced by a second generation PMTs Hamamatsu R6233MOD, 
with higher quantum efficiency (Q.E.) and with lower background with respect 
to those used in phase1; they were produced  after 
a dedicated R\&D in the company, and tests and selections \cite{pmts,ULBNaI}. The new PMTs have Q.E. in 
the range 33-39\% at 420 nm, wavelength of NaI(Tl) emission, and in the range 36-44\% at peak.
The commissioning of the experiment was successfully performed in 2011, allowing the achievement 
of the software energy threshold at 1 keV, and the improvement of some detector's features 
such as energy resolution and acceptance efficiency near software energy threshold 
\cite{pmts}. 

The adopted procedure for noise rejection near software energy threshold 
is discussed in several papers by DAMA collaboration along the years 
and data releases; 
in particular, as regards the data collected in the DAMA/LIBRA--phase2 configuration 
a dedicated discussion is  
presented in section 7 of Ref. \cite{pmts}. The procedure and, in particular, the 
acceptance windows are the same unchanged -- as described there -- along all the 
DAMA/LIBRA--phase2 data taking, 
throughout the months and the annual cycles. The typical behaviour of the overall efficiency 
for {\it single-hit} events as a function of the energy is also shown in section 7 of 
Ref. \cite{pmts}, while in Ref. \cite{bled18} the percentage variations of the efficiency 
are shown, 
considering  all the DAMA/LIBRA--phase2 annual cycles presented here; 
they follow a gaussian distribution
with $\sigma$ = 0.3\% and do not show any modulation with period
and phase as expected for the DM signal \cite{bled18}.


The investigation of the DM annual modulation at lower energy threshold with respect to DAMA/LIBRA--phase1
has been deeply supported by the interest in studying the nature of the DM candidate particles, the features of related astrophysical, 
nuclear and particle physics aspects and by the potentiality of an improved sensitivity in future to investigate both DM annual and diurnal signatures.
Detailed studies will be presented in following papers.

At the end of 2012 new preamplifiers 
and special developed trigger modules were installed and the apparatus was equipped 
with more compact electronic modules \cite{ele_issue}.
Here we just remind that    
the sensitive part of DAMA/LIBRA--phase2 set-up is made of 25 highly radio-pure NaI(Tl) crystal scintillators
(5-rows by 5-columns matrix) having 9.70 kg mass each one;
quantitative analyses of residual contaminants are given in Ref. \cite{perflibra}.
In each detector two 10 cm long UV light guides (made of Suprasil B quartz) act also as
optical windows on the two end faces of the crystal, and are coupled to two low background
PMTs working in coincidence at single photoelectron level. 
The detectors are housed in a sealed low-radioactive
copper box installed in the center of a low-radioactive Cu/Pb/Cd-foils/polyethylene/paraffin shield;
moreover, about 1 m concrete (made from the Gran Sasso rock material) almost fully surrounds (mostly
outside the barrack) this passive shield, acting as a further neutron moderator.
The shield is decoupled from the ground by a metallic structure mounted above a concrete basement;
a neoprene layer separates the concrete basement and the floor of the laboratory. The space between 
this basement and the metallic structure is filled by paraffin for several tens cm in height. 

A threefold-level sealing system prevents the detectors from contact with the environmental air of the 
underground laboratory
and continuously maintains them in HP (high-purity) Nitrogen atmosphere. 
The whole installation is under air conditioning to ensure a suitable and stable working temperature. 
The huge heat capacity of the multi-tons passive shield ($\approx 10^6$ cal/$^o$C) guarantees further relevant 
stability of the detectors' operating temperature. In particular, two independent systems of air 
conditioning are available for redundancy: one cooled by water refrigerated by a dedicated chiller 
and the other operating with cooling gas.
A hardware/software monitoring system provides data on the operating conditions. In particular, 
several probes are read out and the results are stored with the production data. 
Moreover, self-controlled computer based processes automatically monitor several parameters, including 
those from DAQ, and manage the alarms system.
All these procedures, already experienced during DAMA/LIBRA--phase1 \cite{perflibra,modlibra,modlibra2,modlibra3,review}, 
allow us to control and to maintain the running 
conditions stable at a level better than 1\% also in DAMA/LIBRA--phase2 (see e.g. 
Ref. \cite{bled18}).

The light response of the detectors during phase2 typically ranges 
from 6 to 10 photoelectrons/keV, depending on the detector. 
Energy calibration with X-rays/$\gamma$ sources are regularly carried out in the
same running condition down to few keV (for details see e.g. Ref. \cite{perflibra}; in particular, 
double coincidences due to internal X-rays from $^{40}$K 
(which is at ppt levels in the crystals) provide (when summing the data over long periods)
a calibration point at 3.2 keV close to the software energy threshold. 
The DAQ system records both {\it single-hit} events (where just one of the detectors fires) and 
{\it multiple-hit} events (where more than one detector fires) 
up to the MeV region despite the optimization is performed for the lowest energy. 

The radio-purity and details are discussed e.g. in 
Refs. \cite{perflibra,modlibra,modlibra2,modlibra3,review,ULBNaI} and references therein.
The adopted procedures provide sensitivity to large and low mass DM candidates 
inducing nuclear recoils and/or electromagnetic signals.


The data of the former DAMA/NaI setup and, later, those of the DAMA/LIBRA-phase1 have already given (with high confidence level) 
positive evidence for the presence of a signal that satisfies all the requirements of the exploited 
DM annual modulation signature
\cite{modlibra,modlibra2,modlibra3,review,RNC,ijmd}. Moreover, no systematic or side processes able to simultaneously 
satisfy all the many peculiarities of the signature and to account for the whole measured modulation amplitude has been
found or suggested by anyone throughout some decades thus far.

\vspace{0.1cm}
In this paper the model independent result of six annual cycles of DAMA/LIBRA--phase2 is presented. 
The total exposure of DAMA/LIBRA--phase2 is: 
1.13 ton $\times$ yr with an energy threshold at 1 keV; when including also that of the first 
generation DAMA/NaI experiment and DAMA/LIBRA--phase1 the cumulative exposure is
2.46 ton $\times$ yr, corresponding to twenty independent annual cycles.

\section{The DAMA/LIBRA--phase2 annual cycles}

The details of the annual cycles of DAMA/LIBRA--phase2 are reported in Table \ref{tb:years}.
The first annual cycle was dedicated to the commissioning and to the optimizations towards the 
achievement of the 1 keV software energy threshold \cite{pmts}. This period has: 
\begin{table}[!ht]
\caption{Details about the annual cycles of DAMA/LIBRA--phase2. 
The mean value of the squared cosine is $\alpha=\langle cos^2\omega (t-t_0) \rangle$
and the mean value of the cosine is $\beta=\langle cos \omega (t-t_0) \rangle$
(the averages are taken over the live time of the data taking
and $t_0=152.5$ day, i.e.~June 2$^{nd}$);
thus, the variance of the cosine, $(\alpha - \beta^2)$, 
is $\simeq 0.5$ for a detector being operational evenly throughout the year.}
\begin{center}
\vspace{-0.4cm}
\resizebox{\textwidth}{!}{
\begin{tabular}{|c|l|c|c|c|}
\hline
 DAMA/LIBRA--phase2 & Period & Mass (kg) & Exposure (kg$\times$day)  & $(\alpha - \beta^2)$ \\
annual cycle & & & & \\
\hline
 & & \multicolumn{3}{|c|}{ } \\
 1 & Dec.  23, 2010 -- Sept.  9, 2011 & \multicolumn{3}{|c|}{commissioning of phase2} \\
   & & \multicolumn{3}{|c|}{  } \\
 2 & Nov.   2, 2011 -- Sept. 11, 2012 & 242.5 &  62917  & 0.519 \\ 
   &                                  &       &         &       \\ 
 3 & Oct.   8, 2012 -- Sept.  2, 2013 & 242.5 &  60586  & 0.534 \\ 
   &                                  &       &         &       \\ 
 4 & Sept.  8, 2013 -- Sept.  1, 2014 & 242.5 &  73792  & 0.479 \\ 
   &                                  &       &         &       \\ 
 5 & Sept.  1, 2014 -- Sept.  9, 2015 & 242.5 &  71180  & 0.486 \\ 
   &                                  &       &         &       \\ 
 6 & Sept. 10, 2015 -- Aug.  24, 2016 & 242.5 &  67527  & 0.522 \\ 
   &                                  &       &         &       \\ 
 7 & Sept.  7, 2016 -- Sept. 25, 2017 & 242.5 &  75135  & 0.480 \\
   &                                  &       &         &       \\ 
\hline
 DAMA/LIBRA--phase2 & Nov. 2, 2011 -- Sept. 25, 2017 & \multicolumn{2}{|c|}{411137 $\simeq$ 1.13 ton$\times$yr} & 0.502  \\
\hline
\multicolumn{3}{|l}{DAMA/NaI + DAMA/LIBRA--phase1 + DAMA/LIBRA--phase2:} & \multicolumn{2}{c|}{2.46 ton$\times$yr}  \\
\hline
\hline
\end{tabular}
\label{tb:years}}
\vspace{-0.6cm}
\end{center}
\end{table}
i) no data before/near Dec. 2, 2010; 
ii) data sets with some set-up modifications; 
iii) $(\alpha - \beta^2) = 0.355$ well different from 0.5
(i.e. the detectors were not being operational evenly throughout the year).
Thus, this period cannot be used for the annual modulation studies; however, it has been used for other purposes \cite{pmts,IPP}.  
Therefore, as shown in Table \ref{tb:years} the considered annual cycles of DAMA/LIBRA--phase2 
are six 
(exposure of 1.13 ton$\times$yr).
The cumulative exposure, also considering the former DAMA/NaI and DAMA/LIBRA--phase1, is 2.46 ton$\times$yr.

The total number of events collected for the energy calibrations during DAMA/LIBRA--phase2 is 
about $1.3 \times 10^8$, while about $3.4 \times 10^6$ events/keV have been collected for 
the evaluation of the acceptance window efficiency for noise rejection near the software energy 
threshold \cite{perflibra,pmts}.

As it can be inferred from Table \ref{tb:years}, the duty cycle of the experiment is high, ranging between 76\% and 85\%.
The routine calibrations and, in particular, the data collection for the acceptance windows 
efficiency mainly affect it. 

\begin{figure}[!h]
\begin{center} 
\vspace{-0.5cm}
\includegraphics[width=\textwidth] {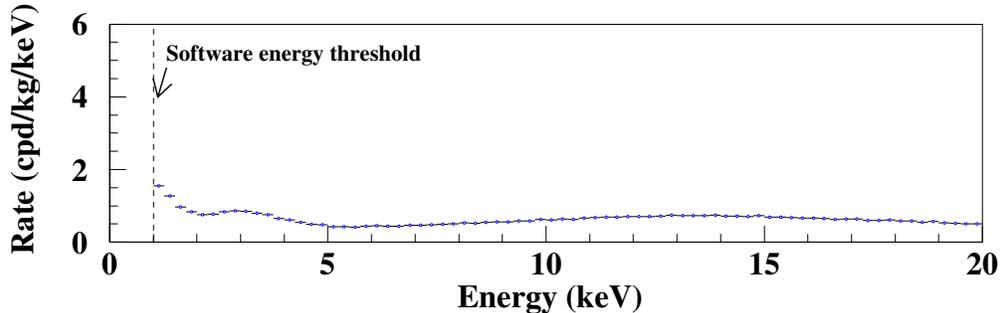}
\end{center}   
\vspace{-0.8cm}
\caption{Cumulative low-energy distribution of the {\it single-hit} scintillation
events (that is each detector has all the others as veto), as measured
by the DAMA/LIBRA--phase2 in an exposure of 1.13 ton $\times$ yr.}
\label{fg:dist}
\end{figure}

Finally, Fig. \ref{fg:dist} shows the low energy distribution of the DAMA/LIBRA--phase2 {\it single-hit} scintillation events. 
It is worth noting that, while DAMA/LIBRA--phase1 showed
a very good linearity between the calibration with the 59.5 keV line of $^{241}$Am and the tagged 3.2 keV line 
of $^{40}$K \cite{perflibra}, in DAMA/LIBRA--phase2 a slight non-linearity is observed 
(it gives a shift of about 0.2 keV at the software energy threshold and vanishes above 15 keV). 
This is taken into account in Fig. \ref{fg:dist} and following analyses 
\footnote{Similar non-linear effects cannot be highlighted in experiments where the energy scale is extrapolated from
calibrations to much higher energies or estimated through MonteCarlo modeling.}.


\section{The annual modulation of the residual rate}

The same procedures already adopted for the DAMA/LIBRA--phase1 \cite{perflibra,modlibra,modlibra2,modlibra3,review} 
have been exploited in the analysis of DAMA/LIBRA--phase2.

\begin{figure}[!ht]
\begin{center}
\vspace{-0.5cm}
\includegraphics[width=\textwidth] {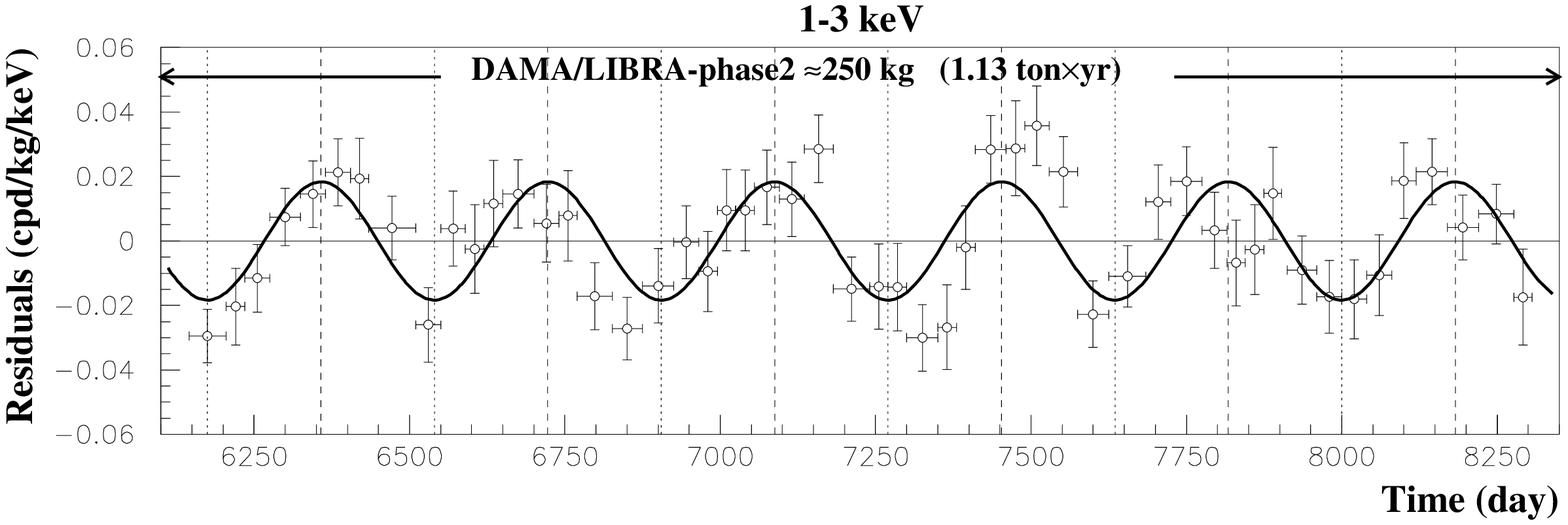}
\includegraphics[width=\textwidth] {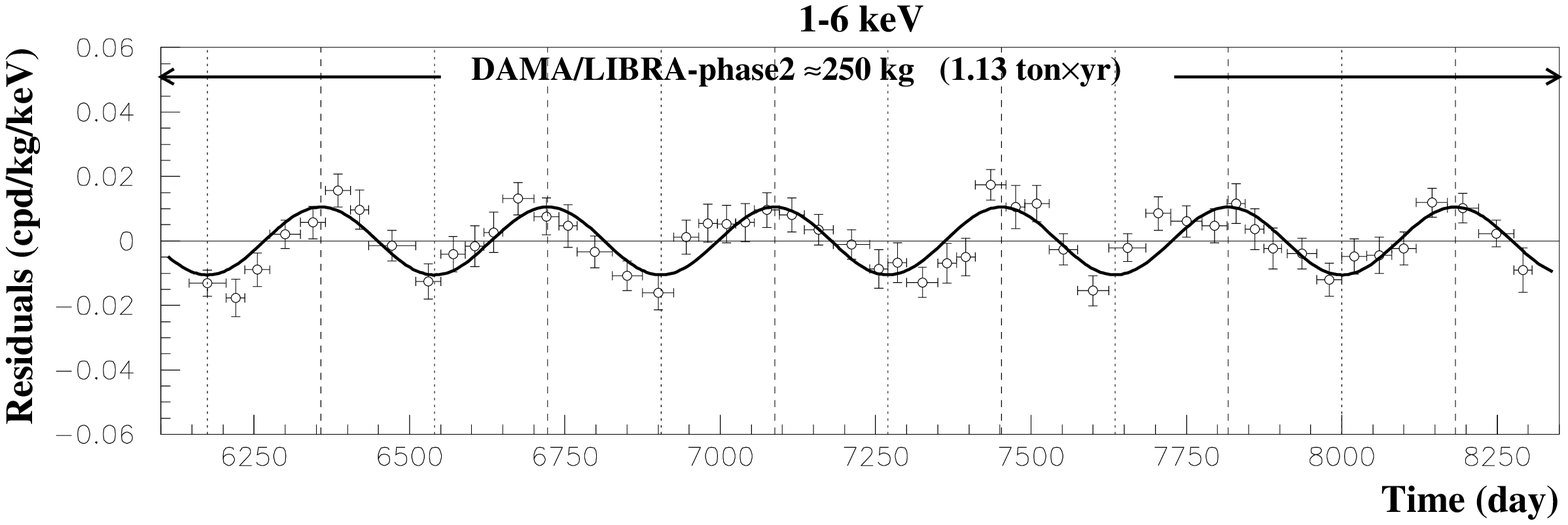} 
\end{center}
\vspace{-0.8cm}
\caption{Experimental residual rate of the {\it single-hit} scintillation events
measured by DAMA/LIBRA--phase2 in the (1--3), (1--6) keV energy intervals
as a function of the time. The time scale is maintained the same of the previous DAMA papers
for consistency.
The data points present the experimental errors as vertical bars and the associated
time bin width as horizontal bars.
The superimposed curves are the cosinusoidal functional forms $A \cos \omega(t-t_0)$
with a period $T = \frac{2\pi}{\omega} =  1$ yr, a phase $t_0 = 152.5$ day (June 2$^{nd}$) and
modulation amplitudes, $A$, equal to the central values obtained by best fit 
on the data points of the entire DAMA/LIBRA--phase2.
The dashed vertical lines
correspond to the maximum expected for the DM signal (June 2$^{nd}$), while
the dotted vertical lines correspond to the minimum.}
\label{fg:res}
\end{figure}

\vspace{0.3cm}

Fig.~\ref{fg:res} shows the time behaviour of the experimental 
residual rates of the {\it single-hit} scintillation 
events in the (1--3), and (1--6) keV energy intervals for the DAMA/LIBRA--phase2 period. 
The residual rates are calculated from the measured rate of the {\it single-hit} events 
after subtracting the constant part, as described in Refs. \cite{modlibra,modlibra2,modlibra3,review,RNC,ijmd}.
The null modulation hypothesis is rejected at very high C.L. by $\chi^2$ test: 
$\chi^2/d.o.f.$ = 127.3/52 and 150.3/52, respectively. The P-values 
are P = 3.0 $\times$ 10$^{-8}$ and P = 1.7 $\times$ 10$^{-11}$, respectively.
The residuals of the DAMA/NaI data (0.29 ton $\times$ yr) are given 
in Ref.~\cite{modlibra,review,RNC,ijmd}, while those of DAMA/LIBRA--phase1 (1.04 ton $\times$ yr)
in Ref.~\cite{modlibra,modlibra2,modlibra3,review}.

\vspace{0.3cm}

\begin{figure}[!ht]
\begin{center}
\includegraphics[width=\textwidth] {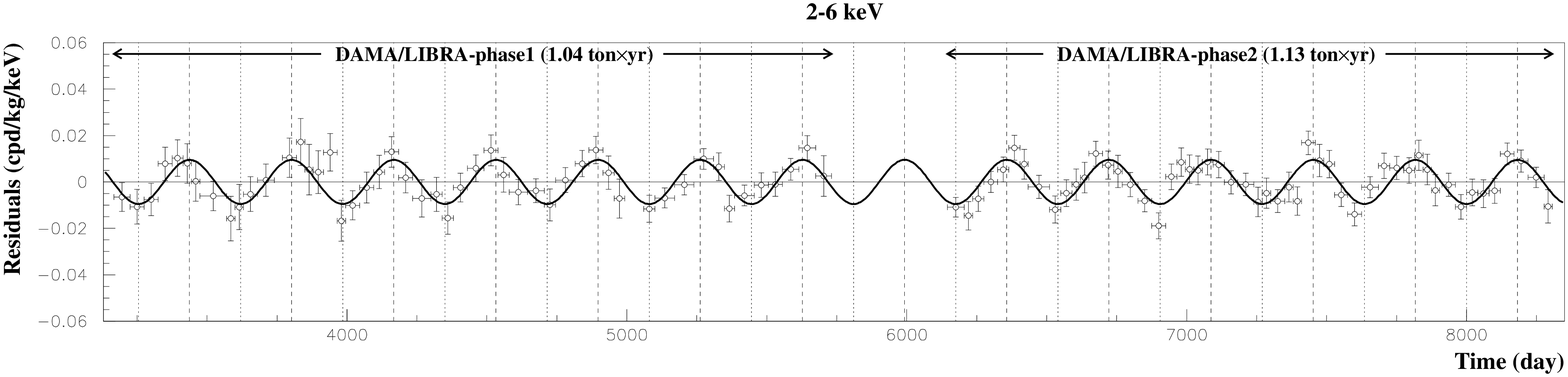}
\end{center}
\vspace{-0.4cm}
\caption{Experimental residual rate of the {\it single-hit} scintillation events
measured by DAMA/LIBRA--phase1 and DAMA/LIBRA--phase2 
in the (2--6) keV energy intervals
as a function of the time. 
The superimposed curve is the cosinusoidal functional forms $A \cos \omega(t-t_0)$
with a period $T = \frac{2\pi}{\omega} =  1$ yr, a phase $t_0 = 152.5$ day (June 2$^{nd}$) and
modulation amplitude, $A$, equal to the central value obtained by best fit 
on the data points of DAMA/LIBRA--phase1 and DAMA/LIBRA--phase2.
For details see Fig. \ref{fg:res}.}
\label{fg:res2}
\end{figure}

The former DAMA/LIBRA--phase1 and the new DAMA/LIBRA--phase2 residual rates of the {\it single-hit} scintillation events 
are reported in Fig.~\ref{fg:res2}. The energy interval is from 2 keV, the software energy threshold of DAMA/LIBRA--phase1,
up to 6 keV.
The null modulation hypothesis is rejected at very high C.L. by $\chi^2$ test: 
$\chi^2/d.o.f.$ = 199.3/102, corresponding to P-value = 2.9 $\times$ 10$^{-8}$.

\vspace{0.3cm}

\begin{table}[!t]
\caption{Modulation amplitude, $A$, obtained by fitting the {\it single-hit} residual rate of 
DAMA/LIBRA--phase2, as reported in Fig.~\ref{fg:res}, and also including 
the residual rates of the former DAMA/NaI and DAMA/LIBRA--phase1.
It was obtained by fitting the data with the formula: 
$A \cos \omega(t-t_0)$.
The period $T = \frac{2\pi}{\omega}$ and the phase $t_0$ are kept fixed at 1 yr and at 152.5
day (June 2$^{nd}$), respectively, as expected by the DM annual modulation signature,
and alternatively kept free.
The results are well compatible with expectations for a signal in the DM annual modulation signature.}
\begin{center}
\begin{tabular}{|rccccc|}
\hline
 &           & $A$ (cpd/kg/keV)& $T = \frac{2\pi}{\omega}$ (yr) &  $t_0$ (days) & C.L. \\
\hline
\hline
\multicolumn{6}{|l|}{DAMA/LIBRA--phase2:} \\
\hspace{1.5cm}
 &  1-3 keV   &   (0.0184$\pm$0.0023)  & 1.0                 & 152.5        &  8.0 $\sigma$  \\
 &  1-6 keV   &   (0.0105$\pm$0.0011)  & 1.0                 & 152.5        &  9.5 $\sigma$  \\
 &  2-6 keV   &   (0.0095$\pm$0.0011)  & 1.0                 & 152.5        &  8.6 $\sigma$  \\
\cline{2-6}
 &  1-3 keV   &   (0.0184$\pm$0.0023)  & (1.0000$\pm$0.0010) & 153$\pm$7    &  8.0 $\sigma$  \\
 &  1-6 keV   &   (0.0106$\pm$0.0011)  & (0.9993$\pm$0.0008) & 148$\pm$6    &  9.6 $\sigma$  \\
 &  2-6 keV   &   (0.0096$\pm$0.0011)  & (0.9989$\pm$0.0010) & 145$\pm$7    &  8.7 $\sigma$  \\
\hline
\hline
\multicolumn{6}{|l|}{DAMA/LIBRA--phase1 + phase2:} \\
 &  2-6 keV   &   (0.0095$\pm$0.0008)  & 1.0                 & 152.5        &  11.9 $\sigma$ \\
\cline{2-6}
 &  2-6 keV   &   (0.0096$\pm$0.0008)  & (0.9987$\pm$0.0008) & 145$\pm$5    &  12.0 $\sigma$ \\
\hline
\hline
\multicolumn{6}{|l|}{DAMA/NaI + DAMA/LIBRA--phase1 + phase2:} \\
 &  2-6 keV   &   (0.0102$\pm$0.0008)  & 1.0                 & 152.5        &  12.8 $\sigma$ \\
\cline{2-6}
 &  2-6 keV   &   (0.0103$\pm$0.0008)  & (0.9987$\pm$0.0008) & 145$\pm$5    &  12.9 $\sigma$ \\
\hline
\hline
\end{tabular}
\end{center}
\label{tb:amp_tot}
\end{table}

\vspace{0.3cm}

The {\it single-hit} residual rates of the DAMA/LIBRA--phase2 (Fig.~\ref{fg:res}) have been fitted 
with the function: $A \cos \omega(t-t_0)$, considering a
period $T = \frac{2\pi}{\omega} =  1$ yr and a phase $t_0 = 152.5$ day (June 2$^{nd}$) as 
expected by the DM annual modulation signature; this can be repeated for the only case of (2-6) keV energy interval
also including the former DAMA/NaI and DAMA/LIBRA--phase1 data.
The goodness of the fits is well supported by the $\chi^2$ test; for example, 
$\chi^2/d.o.f. = 61.3/51, 50.0/51, 113.8/138$ are obtained 
for the (1--3) keV and (1--6) keV cases of DAMA/LIBRA--phase2, and 
for the (2--6) keV case of DAMA/NaI, DAMA/LIBRA--phase1 and DAMA/LIBRA--phase2, respectively.
The results of the best fits are summarized in Table \ref{tb:amp_tot}.
Table \ref{tb:amp_tot} also shows the results of the fit obtained for DAMA/LIBRA--phase2 either including or not DAMA/NaI
and DAMA/LIBRA--phase1, when the period and the phase are kept free in the fitting procedure. As reported in the table,
the period and the phase are well compatible with expectations for 
a DM annual modulation signal. In particular, the phase is consistent 
with about June $2^{nd}$ and is fully consistent with the value independently determined by Maximum Likelihood 
analysis (see later).
For completeness, we recall that a slight energy dependence of the phase 
could be expected (see e.g. Refs. \cite{Fre04_1,Fre04_2,epj06,Fre05,Gel01,caus}), providing intriguing information
on the nature of Dark Matter candidate and related aspects.

\section{Absence of modulation of the background}

Careful investigations on absence of any systematics or side reaction able 
to account for the measured modulation amplitude and to simultaneously 
satisfy all the requirements of the signature have been quantitatively 
carried out also in the past 
(see e.g. Refs. \cite{review}, 
and references therein); none is available. In particular, the cases of muons, 
neutrons and neutrinos have also been carefully investigated, as reported 
in Refs.  \cite{mu,norole}.

As done in previous data releases, absence of any significant 
background modulation in the energy spectrum has also been verified 
in the present data taking for energy regions not of interest for DM.
In fact, the background in the lowest energy region is
essentially due to ``Compton'' electrons, X-rays and/or Auger
electrons, muon induced events, etc., which are strictly correlated
with the events in the higher energy region of the spectrum.
Thus, if a modulation detected in the lowest energy region were due to
a modulation of the background (rather than to a signal),
an equal or larger modulation in the higher energy regions should be present.
\begin{figure}[!h]
\begin{center}
\vspace{-0.4cm}
\includegraphics[width=5.0cm] {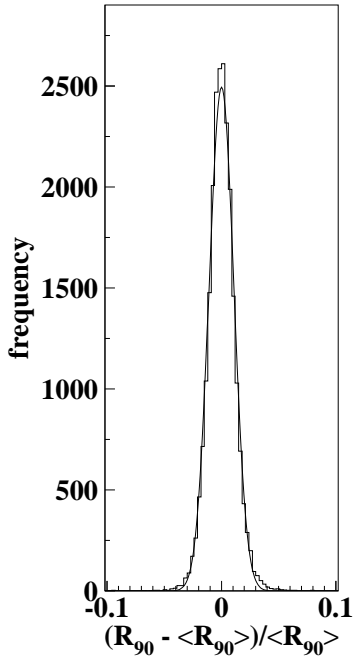}
\end{center}
\vspace{-0.5cm}
\caption{Distribution of the percentage variations
of R$_{90}$ with respect to the mean values for all the detectors in the DAMA/LIBRA--phase2
(histogram); the superimposed curve is a gaussian fit.}
\label{fig_r90}
\end{figure}

For example, the measured rate
integrated above 90 keV, R$_{90}$, as a function 
of the time has been analysed. Fig.~\ref{fig_r90} 
shows the distribution of the percentage variations
of R$_{90}$ with respect to the mean values for all the detectors
in DAMA/LIBRA--phase2.
It shows a cumulative gaussian behaviour
with $\sigma$ $\simeq$ 1\%, well accounted for by the statistical
spread expected from the used sampling time.

Moreover, fitting the time behaviour of R$_{90}$ including a term
with phase and period as for DM particles, a modulation amplitude $A_{R_{90}}$ compatible with zero
has been found for all the annual cycles (see Table \ref{tb:r90}).
This also excludes the presence of any background
modulation in the whole energy spectrum at a level much
lower than the effect found in the lowest energy region for the {\it single-hit} scintillation events.
In fact, otherwise -- considering the R$_{90}$ mean values --
a modulation amplitude of order of tens cpd/kg would be present for each annual cycle,
that is $\simeq$ 100 $\sigma$ far away from the measured values.
\begin{table}[!ht]
\caption{
Modulation amplitudes, $A_{R_{90}}$, (second column)
obtained by fitting the time behaviour of R$_{90}$
for the six annual cycles of DAMA/LIBRA--phase2, including a term with a cosine function
having phase and period as expected for a DM signal. The obtained amplitudes
are compatible with zero, and incompatible ($\simeq$ 100 $\sigma$)
with modulation amplitudes of tens cpd/kg.
Modulation amplitudes, $A_{(6-14)}$, (third column)
obtained by fitting the time behaviour of the 
residual rates of the {\it single-hit} scintillation events in the 
(6--14) keV energy interval. In the fit the phase and the period are at the values expected for a DM signal. 
The obtained amplitudes are compatible with zero.
}
\begin{center}
\begin{tabular}{|c|c|c|}
\hline
 DAMA/LIBRA--phase2 & $A_{R_{90}}$        & $A_{(6-14)}$  \\
 annual cycle       & (cpd/kg)         & (cpd/kg/keV)    \\
\hline
  2                 &  (0.12$\pm$0.14) &  (0.0032$\pm$0.0017)  \\
  3                 & -(0.08$\pm$0.14) &  (0.0016$\pm$0.0017) \\
  4                 &  (0.07$\pm$0.15) &  (0.0024$\pm$0.0015) \\
  5                 & -(0.05$\pm$0.14) & -(0.0004$\pm$0.0015) \\
  6                 &  (0.03$\pm$0.13) &  (0.0001$\pm$0.0015) \\
  7                 & -(0.09$\pm$0.14) &  (0.0015$\pm$0.0014) \\
\hline
\hline
\end{tabular}
\end{center}
\label{tb:r90}
\end{table}

\begin{figure}[!ht]
\vspace{-0.8cm}
\centering
\includegraphics[width=0.48\textwidth] {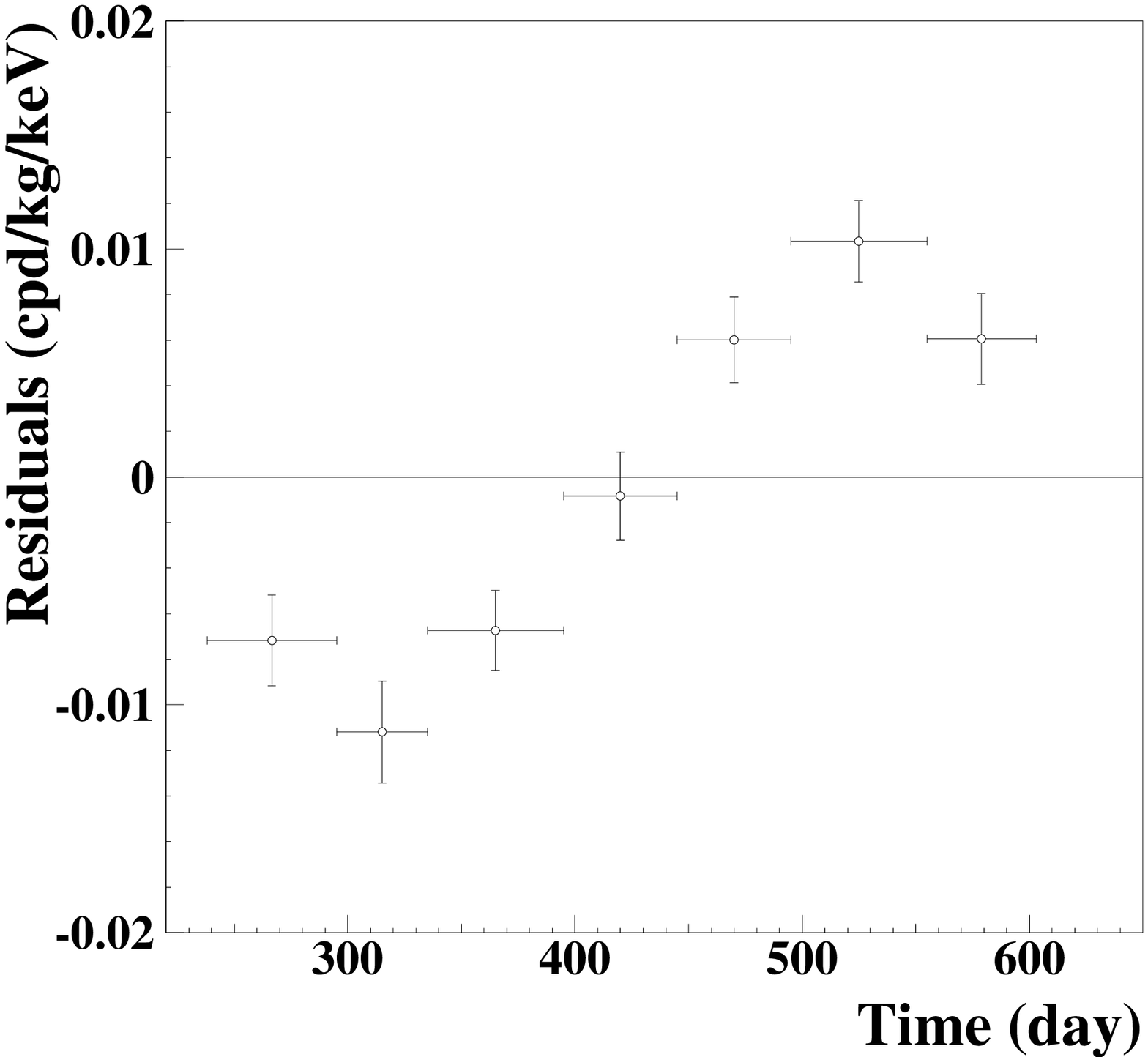}
\includegraphics[width=0.48\textwidth] {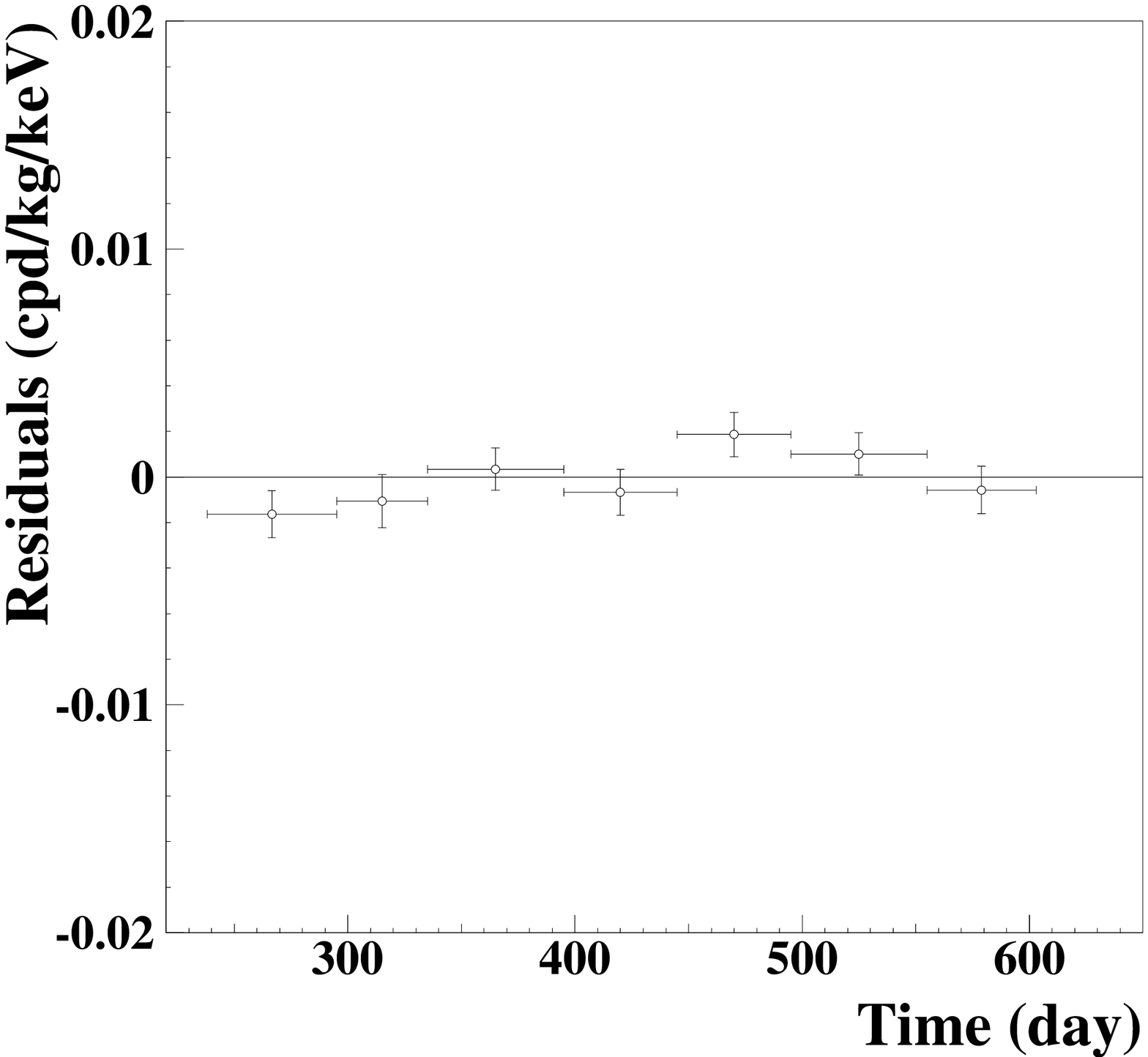}
\vspace{-0.2cm}
\caption{
Experimental {\it single-hit} residuals in the (1--6) keV and in the
(10--20) keV energy regions for DAMA/LIBRA--phase2 as if they were collected in a
single annual cycle (i.e.~binning in the variable time from the January 1st of each annual cycle). 
The data points present the experimental errors 
as vertical bars and the associated
time bin width as horizontal bars. 
The initial time of the figures is taken at August 7$^{th}$.
A clear modulation satisfying all the peculiarities of the 
DM annual modulation signature is present 
in the lowest energy interval with A=(0.0106 $\pm$ 0.0011) cpd/kg/keV,
while it is absent just above: A=(0.0010 $\pm$ 0.0006) cpd/kg/keV.}
\label{fg:res1}
\end{figure}

Similar results are obtained when comparing 
the {\it single-hit} residuals in the (1--6) keV with those 
in other energy intervals; for example Fig.~\ref{fg:res1} shows the 
{\it single-hit} residuals in the (1--6) keV and in the
(10--20) keV energy regions for DAMA/LIBRA--phase2 as if they were collected in a
single annual cycle (i.e.~binning in the variable time from the January 1st of each annual cycle).

Moreover, Table \ref{tb:r90} shows the modulation amplitudes obtained by fitting the time behaviour of the 
residual rates of the {\it single-hit} scintillation events in the 
(6--14) keV energy interval for the DAMA/LIBRA--phase2 annual cycles.
In the fit the phase and the period are at the values expected for a DM signal. 
The obtained amplitudes are compatible with zero.

\begin{figure}[!ht]
\begin{center}
\vspace{-0.8cm}
\includegraphics[width=0.98\textwidth] {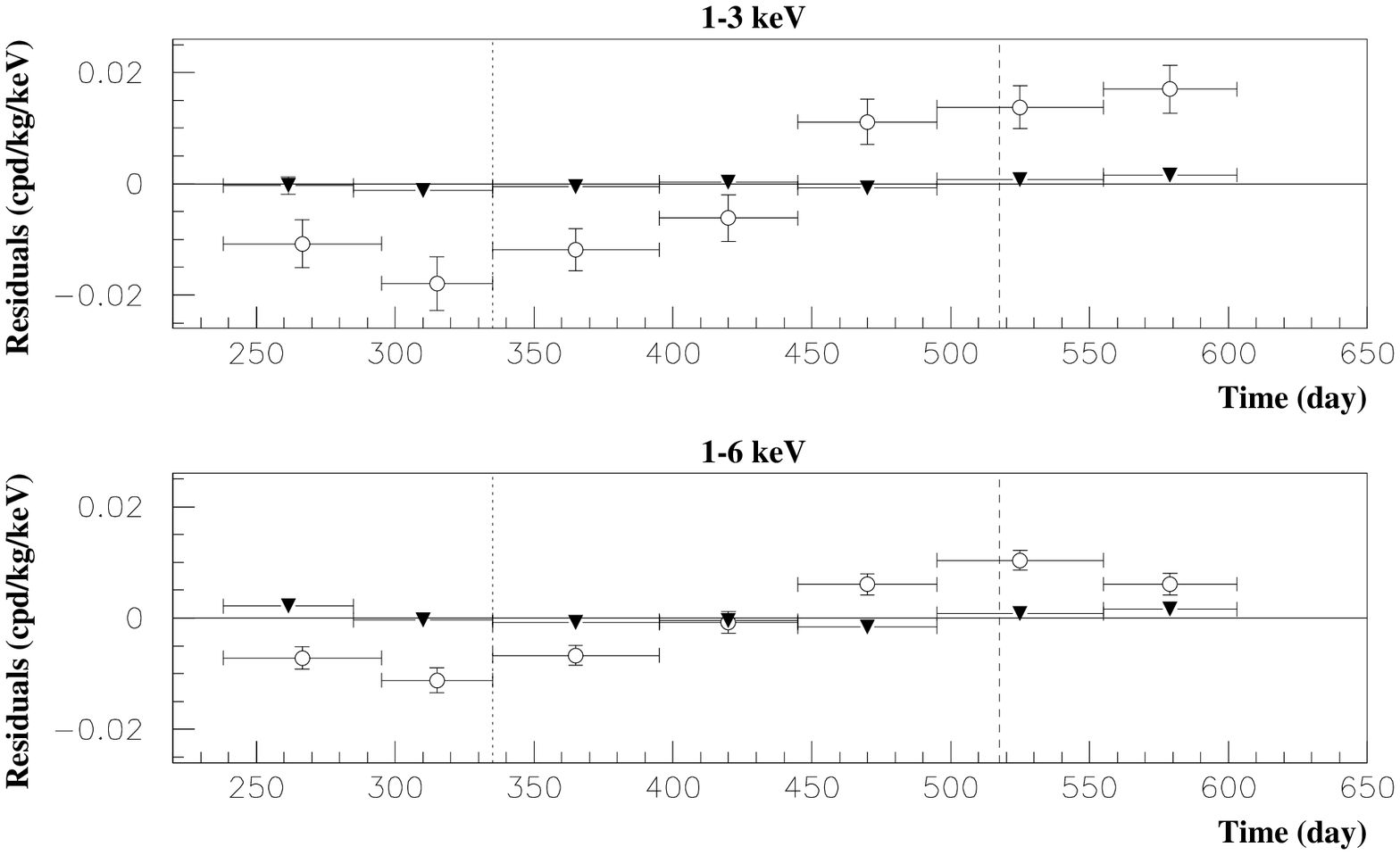}
\end{center}
\vspace{-0.5cm}
\caption{Experimental residual rates of DAMA/LIBRA--phase2 {\it single-hit} events 
(open circles), class of events to which DM events belong, and for {\it multiple-hit} 
events (filled triangles),
class of events to which DM events do not belong.
They have been obtained by considering for each class of events the data as collected in a 
single annual cycle 
and by using in both cases the same identical hardware and the same identical software procedures.
The initial time of the figure is taken on August 7$^{th}$.
The experimental points present the errors as vertical bars 
and the associated time bin width as horizontal 
bars. Analogous results were obtained for DAMA/NaI (two last annual cycles) and DAMA/LIBRA--phase1 
 \cite{modlibra,modlibra2,modlibra3,review,ijmd}.}
\label{fig_mul}
\end{figure}

A further relevant investigation on DAMA/LIBRA--phase2 data has been performed by 
applying the same hardware and software 
procedures, used to acquire and to analyse the {\it single-hit} residual rate, to the 
{\it multiple-hit} one. 
Since the 
probability that a DM particle interacts in more than one detector 
is negligible, a DM signal can be present just in the {\it single-hit} residual rate.
Thus, the comparison of the results of the {\it single-hit} events with those of the  {\it 
multiple-hit} ones corresponds to compare the cases of DM particles beam-on 
and beam-off.
This procedure also allows an additional test of the background behaviour in the same energy interval 
where the positive effect is observed. 

We note that an event is considered multiple when there is a deposition of energy in coincidence in more than one detector of the set-up. The multiplicity 
can, in 
principle, range from 2 to 25.  A multiple event in a given energy interval, say 1-6 keV  is given by an 
energy deposition between 1 and 6 keV in one detector and other deposition(s) in other detector(s). 
The residual rate of events with multiplicity equal or greater than 2 with an energy deposition in the range 1-6 keV is shown in Fig.~\ref{fig_mul};
the only procedure applied to multiple events is that used to reject noise events near software energy threshold and is the same used for sing-hit 
events. 

\vspace{0.2cm}

In particular, in Fig.~\ref{fig_mul} the residual rates of the {\it single-hit} scintillation events collected during DAMA/LIBRA--phase2 
are reported, as collected in a single cycle, together with the residual rates 
of the {\it multiple-hit} events, in the considered energy intervals.
While, as already observed, a clear modulation, satisfying all the peculiarities of the DM
annual modulation signature, is present in 
the {\it single-hit} events,
the fitted modulation amplitudes for the {\it multiple-hit}
residual rate are well compatible with zero:
$  (0.0007\pm0.0006)$ cpd/kg/keV, and
$  (0.0004\pm0.0004)$ cpd/kg/keV,
in the energy regions (1--3) keV, and (1--6) keV, respectively.
Thus, again evidence of annual modulation with proper features as required by the DM annual 
modulation signature is present in the {\it single-hit} residuals (events class to which the
DM particle induced events belong), while it is absent in the {\it multiple-hit} residual 
rate (event class to which only background events belong).
Similar results were also obtained for the two last annual cycles of DAMA/NaI \cite{ijmd}
and for DAMA/LIBRA--phase1 \cite{modlibra,modlibra2,modlibra3,review}.
Since the same identical hardware and the same identical software procedures have been used to 
analyse the two classes of events, the obtained result offers an additional strong support for the 
presence of a DM particle component in the galactic halo.

\vspace{0.2cm}

In conclusion, no background process able to mimic the DM annual modulation
signature (that is, able to simultaneously satisfy 
all the peculiarities of the signature and to account for the measured modulation amplitude)
has been found or suggested by anyone throughout some decades thus far (see also discussions e.g. in 
Ref.~\cite{perflibra,modlibra,modlibra2,modlibra3,review,mu,norole,uni18}).

\section{The analysis in frequency}

\begin{figure}[!hp]
\centering
\vspace{-0.8cm}
\includegraphics[width=0.7\textwidth] {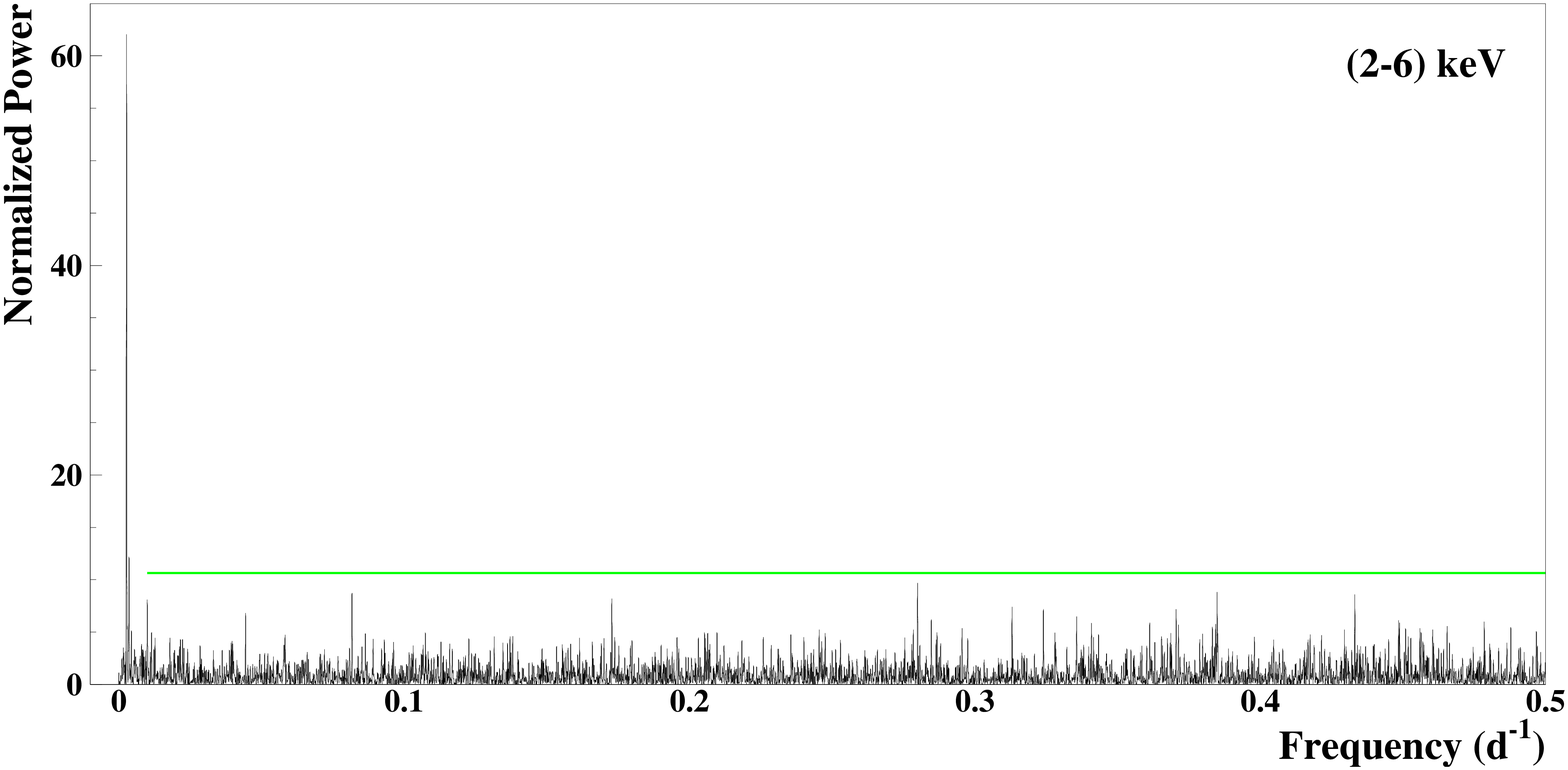}
\includegraphics[width=0.7\textwidth] {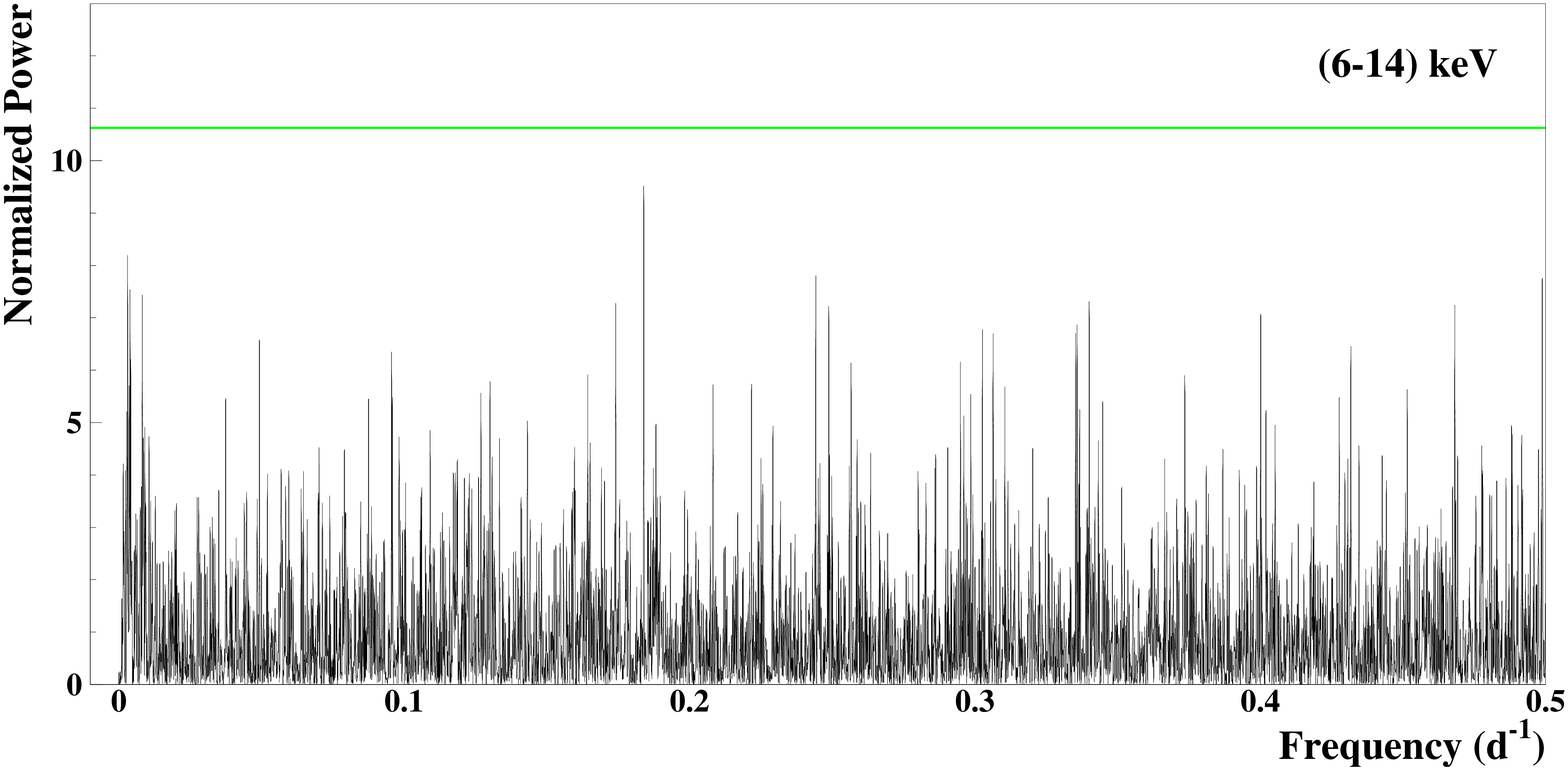}
\includegraphics[width=0.5\textwidth] {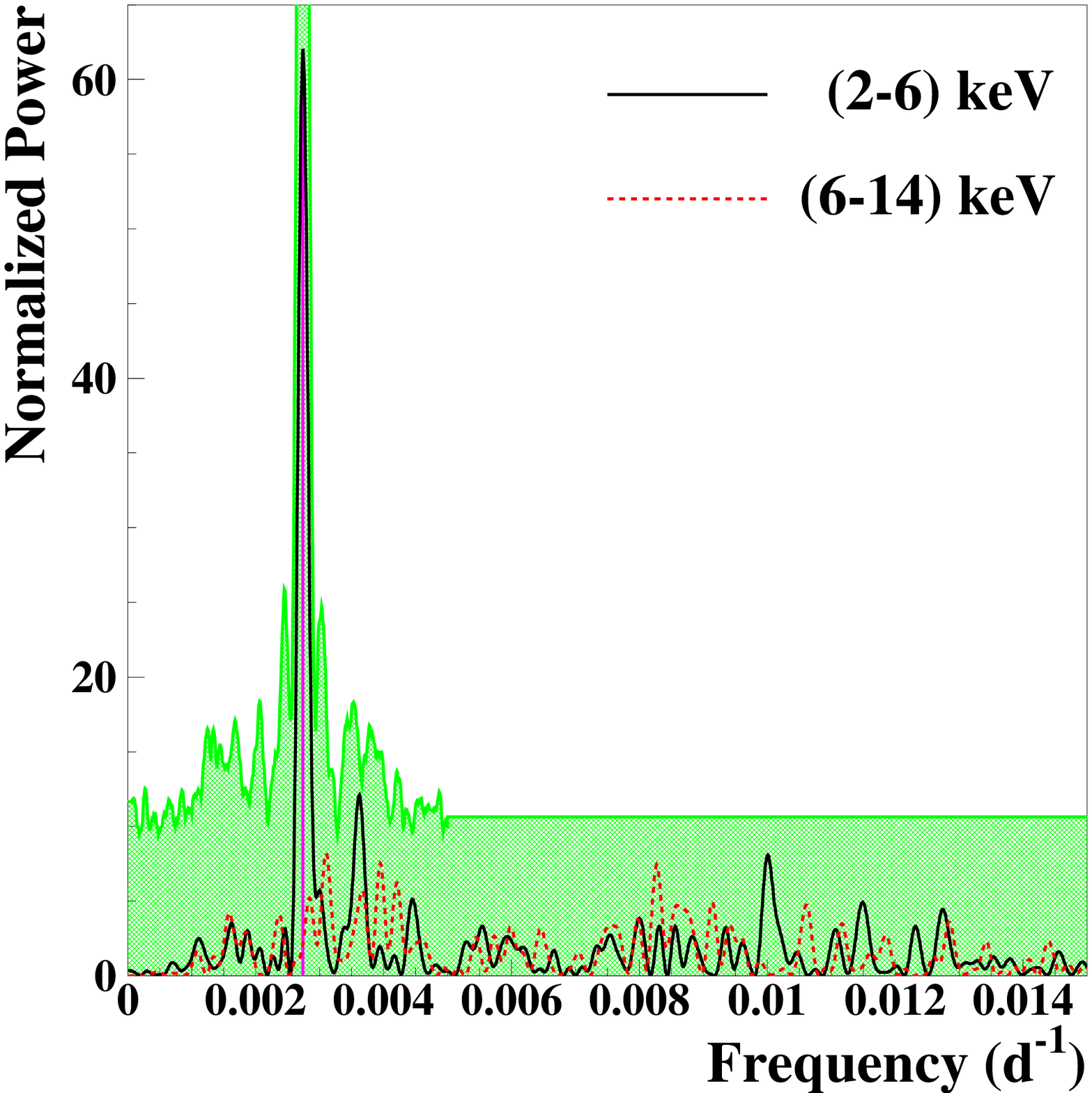}
\vspace{-0.4cm}
\caption{
Power spectra of the time sequence of the measured {\it single-hit} events for DAMA/LIBRA--phase1 and
DAMA/LIBRA--phase2 grouped in 1 day bins.
From top to bottom: spectra up to the Nyquist frequency for (2--6) keV and (6--14) keV energy intervals 
and their zoom around the 1 y$^{-1}$ peak, for (2--6) keV (solid line) and (6--14) keV (dotted line) energy intervals.
The main mode present at the lowest energy interval corresponds to a frequency of $2.74 \times 10^{-3}$ d$^{-1}$
(vertical line, purple on-line). It corresponds to a period of $\simeq$ 1 year.
A similar peak is not present in the (6--14) keV energy interval.
The shaded (green on-line) area in the bottom figure -- calculated by Monte Carlo procedure -- represents the 90\% C.L. 
region where all the peaks are expected to fall for the (2--6) keV energy interval. 
In the frequency range far from the signal for the (2--6) keV energy region and 
for the whole (6--14) keV spectrum, the upper limit of the shaded region (90\% C.L.) can be calculated to be 10.6 (continuous lines, green on-line). 
}
\vspace{-0.3cm}
\label{fg:pwr}
\normalsize
\end{figure}

To perform the Fourier analysis of the DAMA/LIBRA--phase1 and phase2 data in a wider region of considered frequency,
the {\it single-hit} events have been grouped in 1 day bins.
Due to the low statistics in each time bin, a procedure detailed in Ref.~\cite{ranucci} has been followed.
The whole power spectra up to the Nyquist frequency and the zoomed ones are reported in Fig.~\ref{fg:pwr}.
A clear peak corresponding to a period of 1 year is evident for the lowest energy interval;
the same analysis in the (6--14) keV energy region shows only aliasing peaks instead.
Neither other structure at different frequencies has been observed.

\begin{figure}[!hb]
\centering
\vspace{-0.2cm}
\includegraphics[width=0.6\textwidth] {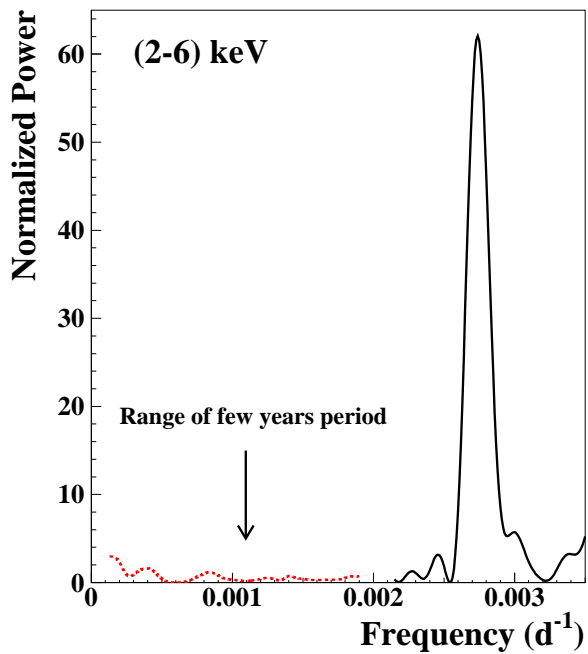}
\vspace{-0.2cm}
\caption{
Power spectrum of the annual baseline counting rates for the {\it single-hit} events of DAMA/LIBRA--phase1 and
DAMA/LIBRA--phase2 in the (2--6) keV energy interval (dotted line, red on-line).
Also shown for comparison is the power spectrum reported in Fig.~\ref{fg:pwr} (solid line).
The calculation has been performed according to Ref. \cite{review}.
As can be seen, a principal mode is present at a frequency of $2.74 \times 10^{-3}$ d$^{-1}$, that corresponds to a
period of $\simeq$ 1 year. No statistically-significant peak is present at lower frequencies.
This implies that no evidence for a long term modulation is present
in the {\it single-hit} scintillation event in the low energy range.}
\label{fg:pwr_bsl}
\normalsize
\end{figure}

As to the significance of the peaks present in the periodogram, we remind that the periodogram ordinate, $z$,
at each frequency follows a simple exponential distribution $e^{-z}$ in the case of the null hypothesis or white
noise \cite{scargle82}.
Therefore, if $M$ independent frequencies are scanned, the probability to obtain values larger than $z$ is:
$P(>z) = 1 - \left( 1 - e^{-z} \right)^M$.

In general $M$ depends on the number of sampled frequencies, the number of data points $N$, and their detailed
spacing. It turns out that $M$ is very nearly equal to $N$ when the data points are approximately equally
spaced, and when the sampled frequencies cover the frequency range from 0 to the Nyquist frequency \cite{press92,horne86}.

\begin{figure}[!ht]
\centering
\vspace{-0.2cm}
\includegraphics[width=0.6\textwidth] {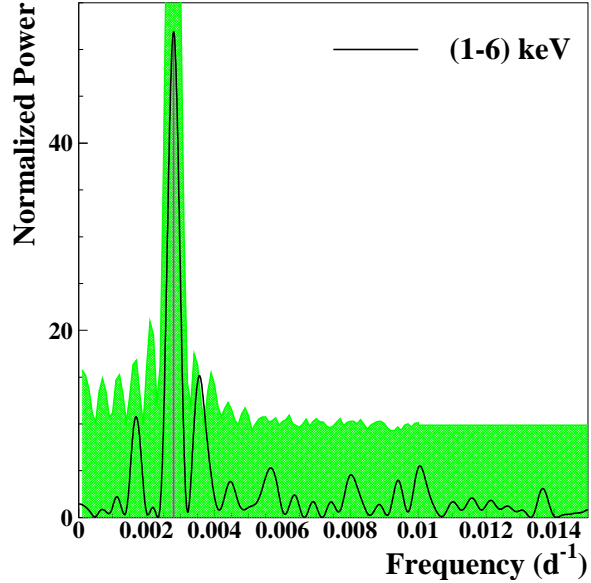}
\vspace{-0.2cm}
\caption{
Power spectrum of the time sequence of the measured {\it single-hit} events in the (1--6) keV energy interval for
DAMA/LIBRA--phase2 grouped in 1 day bin.
The main mode present at the lowest energy interval corresponds to a frequency of $2.79 \times 10^{-3}$ d$^{-1}$ (vertical
line, purple on-line). It corresponds to a period of $\simeq$ 1 year.
The shaded (green on-line) area -- calculated by Monte Carlo procedure -- represents the 90\% C.L. region where all the peaks 
are expected to fall for the (1--6) keV energy interval.}
\label{fg:pwr_16}
\normalsize
\end{figure}

The number of data points used to obtain the spectra in Fig.~\ref{fg:pwr} is $N=4341$ (days measured over the 4748 days of
the 13 DAMA/LIBRA--phase1 and phase2 annual cycles) and the full frequencies region up to Nyquist frequency has been scanned.
Therefore, assuming $M=N$, the significance levels $P=$ 0.10, 0.05 and 0.01, correspond to peaks with heights larger than 
$z=$ 10.6, 11.3 and 13.0, respectively, in the spectra of Fig~\ref{fg:pwr}.

In the case below 6 keV, a signal is present; thus, to properly evaluate the C.L. the signal must be included. 
This has been done by a dedicated Monte Carlo procedure where a large number of similar experiments has been simulated.
The 90\% C.L. region (shaded, green on-line) where all the peaks are expected to fall for the (2--6) keV energy interval is reported in Fig~\ref{fg:pwr}.
Several peaks, satellite of the one year period frequency, are present.

In conclusion, apart from the peak corresponding to a 1 year period, no other peak is statistically significant either in the
low and high energy regions.

Moreover, for each annual cycle of DAMA/LIBRA--phase1 and phase2, the annual baseline counting rates have been calculated for
the (2--6) keV energy interval.
Their power spectrum in the frequency range $0.0002-0.0018$ d$^{-1}$ (corresponding to a period range 13.7--1.5 year) is reported
in Fig.~\ref{fg:pwr_bsl}.
The power spectrum (solid black line) above 0.0022 d$^{-1}$ of Fig.~\ref{fg:pwr} is reported for comparison.
The calculation has been performed according to Ref. \cite{review}.
No statistically-significant peak is present at frequencies lower than 1 y$^{-1}$.
This implies that no evidence for a long term modulation in the counting rate is present.

Finally, the case of the (1--6) keV energy interval of the DAMA/LIBRA--phase2 data is reported in Fig.~\ref{fg:pwr_16}.
As previously the only significant peak is the one corresponding to one year period.
No other peak is statistically significant being below the shaded (green on-line) area obtained by Monte Carlo procedure.

\section{The modulation amplitudes by the maximum likelihood approach}

The annual modulation present at low energy can also 
be pointed out by depicting the energy dependence of
the modulation amplitude, $S_{m}(E)$, obtained
by maximum likelihood method considering fixed period and phase: $T=$1 yr and $t_0=$ 152.5 day.
For such purpose the likelihood function of the {\it single-hit} experimental data
in the $k-$th energy bin is defined as: $ {\it\bf L_k}  = {\bf \Pi}_{ij} e^{-\mu_{ijk}}
{\mu_{ijk}^{N_{ijk}} \over N_{ijk}!}$,
where $N_{ijk}$ is the number of events collected in the
$i$-th time interval (hereafter 1 day), by the $j$-th detector and in the
$k$-th energy bin. $N_{ijk}$ follows a Poisson's
distribution with expectation value
$\mu_{ijk} = \left[ b_{jk} + S_{i}(E_k) \right] M_j \Delta
t_i \Delta E \epsilon_{jk}$.
The $b_{jk}$ are the background contributions, $M_j$ is the mass of the $j-$th detector,
$\Delta t_i$ is the detector running time during the $i$-th time interval,
$\Delta E$ is the chosen energy bin,
$\epsilon_{jk}$ is the overall efficiency. 
The signal can be written
as: 

$$S_{i}(E) = S_{0}(E) + S_{m}(E) \cdot \cos\omega(t_i-t_0),$$ 
where $S_{0}(E)$ is the constant part of 
the signal and $S_{m}(E)$ is the modulation amplitude.
The usual procedure is to minimize the function $y_k=-2ln({\it\bf L_k}) - const$ for each energy bin;
the free parameters of the fit are the $(b_{jk} + S_{0})$ contributions and the $S_{m}$
parameter. 

\begin{figure}[!h]
\begin{center}
\includegraphics[width=\textwidth] {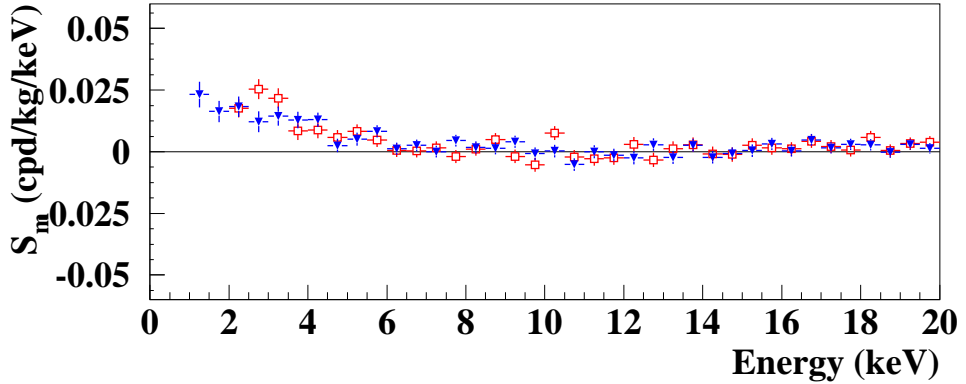}
\end{center}
\vspace{-0.8cm}
\caption{Modulation amplitudes, $S_{m}$, for DAMA/LIBRA--phase2 (exposure 1.13 ton$\times$yr)
from the energy threshold of 1 keV up to 20 keV (full triangles, blue data points on-line) -- and for DAMA/NaI and DAMA/LIBRA--phase1 
(exposure 1.33 ton$\times$yr) \cite{modlibra3} (open squares, red data points on-line). The energy bin $\Delta E$ is 0.5 keV.
The modulation amplitudes obtained in the two data sets are consistent in the (2--20) keV:
the $\chi^2$ is 32.7 for 36 $d.o.f.$, and the corresponding P-value is 63\%.
In the (2--6) keV energy region, where the signal is present, the $\chi^2/d.o.f.$ is 10.7/8 (P-value = 22\%).
}
\label{fg:sme_nailib1_vs_lib2}
\end{figure}

In Fig.~\ref{fg:sme_nailib1_vs_lib2} the modulation amplitudes obtained considering the DAMA/LIBRA--phase2 data
are reported as full triangles (blue points on-line) from the energy threshold of 1 keV up to 20 keV. 
Superimposed to the picture as open squared (red on-line) data points  
are the modulation amplitudes of the 
former DAMA/NaI and DAMA/LIBRA--phase1 \cite{modlibra3}.
The modulation amplitudes obtained in the two data sets are consistent in the (2--20) keV,
since the $\chi^2$ is 32.7 for 36 $d.o.f.$ corresponding to P-value = 63\%.
In the (2--6) keV energy region, where the signal is present, the $\chi^2/d.o.f.$ is 10.7/8 (P-value = 22\%).

\begin{figure}[!h]
\begin{center}
\includegraphics[width=\textwidth] {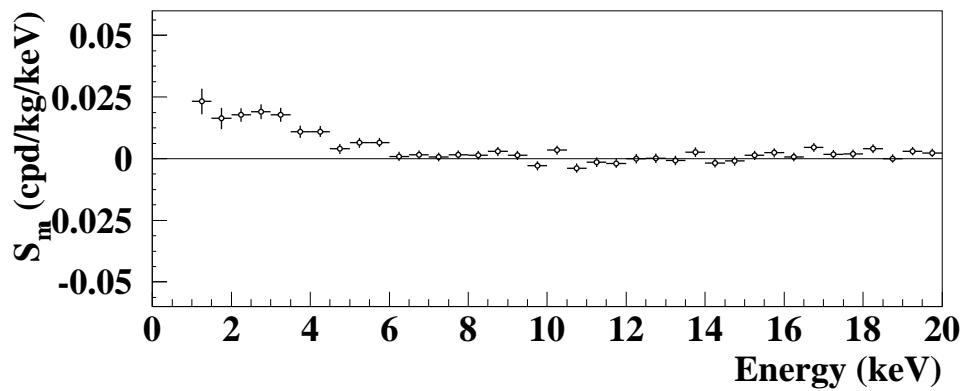}
\end{center}
\vspace{-0.6cm}
\caption{Modulation amplitudes, $S_{m}$, for the whole data sets: DAMA/NaI, DAMA/LIBRA--phase1 and DAMA/LIBRA--phase2 
(total exposure 2.46 ton$\times$yr) above 2 keV; below 2 keV only the DAMA/LIBRA-phase2 exposure (1.13 ton $\times$ yr) is available and used. The 
energy bin $\Delta E$ is 0.5 keV.
A clear modulation is present in the lowest energy region,
while $S_{m}$ values compatible with zero are present just above. In fact, the $S_{m}$ values
in the (6--20) keV energy interval have random fluctuations around zero with
$\chi^2$ equal to 42.6
for 28 $d.o.f.$ (upper tail probability of 4\%).
}
\label{fg:sme}
\end{figure}

\begin{table}[!ht]
	\caption{Modulation amplitudes, $S_{m}$, for the whole data sets: DAMA/NaI, DAMA/LIBRA--phase1 and DAMA/LIBRA--phase2
                (total exposure 2.46 ton$\times$yr); data below 2 keV refer instead only to the DAMA/LIBRA-phase2 exposure (1.13 ton$\times$yr).}
	\begin{center}
		\begin{tabular}{|cc|cc|}
			\hline
			Energy          &   $S_m$ (cpd/kg/keV)   &   Energy          &   $S_m$ (cpd/kg/keV)   \\
			\hline
			(1.0--1.5) keV  &   (0.0232$\pm$0.0052)  &   (6.5--7.0) keV  &   (0.0016$\pm$0.0018)  \\
			(1.5--2.0) keV  &   (0.0164$\pm$0.0043)  &   (7.0--7.5) keV  &   (0.0007$\pm$0.0018)  \\
			(2.0--2.5) keV  &   (0.0178$\pm$0.0028)  &   (7.5--8.0) keV  &   (0.0016$\pm$0.0018)  \\
			(2.5--3.0) keV  &   (0.0190$\pm$0.0029)  &   (8.0--8.5) keV  &   (0.0014$\pm$0.0018)  \\
			(3.0--3.5) keV  &   (0.0178$\pm$0.0028)  &   (8.5--9.0) keV  &   (0.0029$\pm$0.0018)  \\
			(3.5--4.0) keV  &   (0.0109$\pm$0.0025)  &   (9.0--9.5) keV  &   (0.0014$\pm$0.0018)  \\
			(4.0--4.5) keV  &   (0.0110$\pm$0.0022)  &  (9.5--10.0) keV  &  -(0.0029$\pm$0.0019)  \\
			(4.5--5.0) keV  &   (0.0040$\pm$0.0020)  & (10.0--10.5) keV  &   (0.0035$\pm$0.0019)  \\
			(5.0--5.5) keV  &   (0.0065$\pm$0.0020)  & (10.5--11.0) keV  &  -(0.0038$\pm$0.0019)  \\
			(5.5--6.0) keV  &   (0.0066$\pm$0.0019)  & (11.0--11.5) keV  &  -(0.0013$\pm$0.0019)  \\
			(6.0--6.5) keV  &   (0.0009$\pm$0.0018)  & (11.5--12.0) keV  &  -(0.0019$\pm$0.0019)  \\
			\hline
		\end{tabular}
	\end{center}
	\label{tb:sm2}
\end{table}

As shown in Fig.~\ref{fg:sme_nailib1_vs_lib2} positive signal is present below 6 keV 
also in the case of DAMA/LIBRA--phase2. Above 6 keV the $S_{m}$ values are compatible with zero; actually, they
have random fluctuations around zero, since the $\chi^2$ in the (6--20) keV energy interval for the DAMA/LIBRA--phase2
data is equal to 29.8 for 28 $d.o.f.$ (upper tail probability of 37\%).
Similar considerations have been done for DAMA/NaI and DAMA/LIBRA--phase1 where the $\chi^2$ 
in the (6--20) keV energy interval is 35.8 for 28 $d.o.f.$ (upper tail probability of 15\%) \cite{modlibra3}.

The modulation amplitudes for the whole data sets: DAMA/NaI, DAMA/LIBRA--phase1 and DAMA/LIBRA--phase2 
(total exposure 2.46 ton$\times$yr) plotted in Fig.~\ref{fg:sme};  
the data below 2 keV refer only to the 
DAMA/LIBRA-phase2 exposure (1.13 ton$\times$yr).
It can be inferred that positive signal is present in the (1--6) keV energy interval, while $S_{m}$
values compatible with zero are present just above. 
All this confirms the previous analyses. 
In Table \ref{tb:sm2} the values of the modulation amplitudes of the (1--12) keV energy region are also reported. 
The test of the hypothesis that the $S_{m}$ values
in the (6--14) keV energy interval have random fluctuations around zero 
yields $\chi^2$ equal to 19.0 for 16 $d.o.f.$ (upper tail probability of 27\%).

For the case of (6--20) keV energy interval $\chi^2/d.o.f.=$ 42.6/28 (upper tail probability of 4\%).
The obtained $\chi^2$ value is rather large due mainly to two data points, whose centroids are at 16.75 and 18.25 keV,
far away from the (1--6) keV energy interval.
The P-values obtained by excluding only the first and either the points are 11\% and 25\%.


It worth noting that in the DAMA experiments the exploited DM model-independent annual modulation signature does not require any 
identification of the constant part of the signal $S_0$ from the {\it single-hit} counting rate, in order to establish 
the presence of a signal ($S_m$); 
in fact, the modulation amplitudes, $S_m$, are the experimental observables. No background subtraction is applied since the exploited 
signature itself acts as an 
effective background rejection, as pointed out since the early papers by Freese et al.\footnote{Anyhow, the $S_m/S_0$ ratio is of 
interest in the corollary model dependent analyses in the framework of specific astrophysical, nuclear and particle physics scenarios 
(not discussed in the present paper). Thus, exploiting a simple and safe approach, 
the lower limit on the $S_m/S_0$ ratio has been given for DAMA/LIBRA--phase1 e.g. in Ref. \cite{mirasim,bled14}.
In DAMA/LIBRA--phase2 the upper limit on $S_0$ is estimated with the same procedure to be about 0.80 cpd/kg/keV, and 0.24 cpd/kg/keV, 
in the (1--2) keV and (2--3) keV energy intervals,
corresponding to the $S_m/S_0$ ratio $\gsim 2.4\%$, and $\gsim 6.3\%$, respectively.}.


\subsection{The $S_m$ distributions}

\begin{figure}[!ht]
\vspace{-.7cm}
\begin{center}
\includegraphics[width=0.65\textwidth] {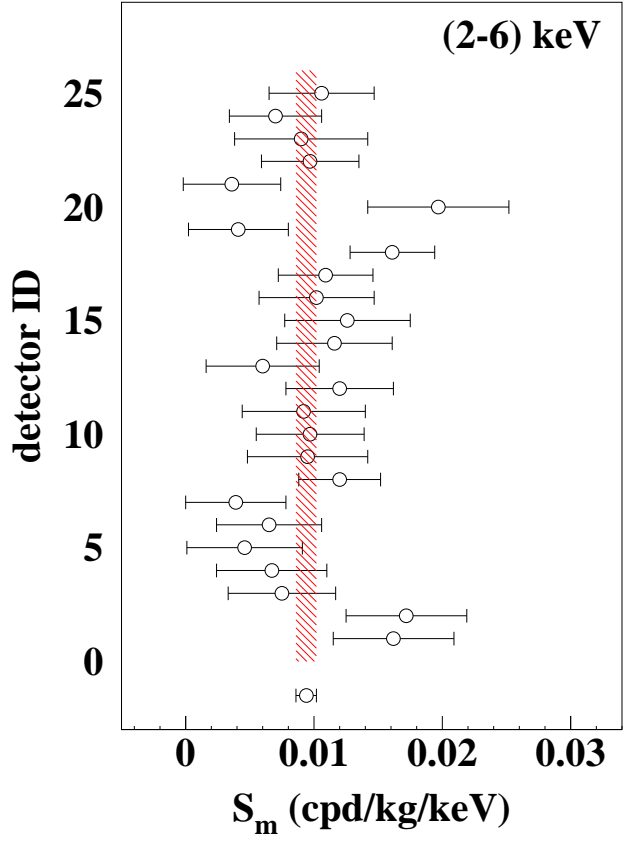}
\end{center}
\vspace{-.6cm}
\caption{Modulation amplitudes $S_m$ integrated in the range (2--6) keV for each of the 25 detectors 
for the DAMA/LIBRA--phase1 and DAMA/LIBRA--phase2 periods.
The errors are at $1\sigma$ confidence level. The weighted averaged point and $1\sigma$ band (shaded area) are also reported. 
The $\chi^2$ is 23.9 over 24 $d.o.f.$, supporting the hypothesis that 
the signal is well distributed over all the 25 detectors.
}
\label{fg:democ}
\end{figure}

The method also allows the extraction of the $S_{m}$ values for each detector. In particular,
the modulation amplitudes $S_m$ integrated in the range (2--6) keV for each of the 25 detectors 
for the DAMA/LIBRA--phase1 and DAMA/LIBRA--phase2 periods are reported in Fig.~\ref{fg:democ}.
They have random fluctuations around the weighted averaged value (shaded band) confirmed by the 
$\chi^2/d.o.f.$ equal to 23.9/24. Thus, the hypothesis that 
the signal is well distributed over all the 25 detectors is accepted.

\begin{figure}[!ht]
\vspace{-1.2cm}
\begin{center}
\includegraphics[width=0.7\textwidth] {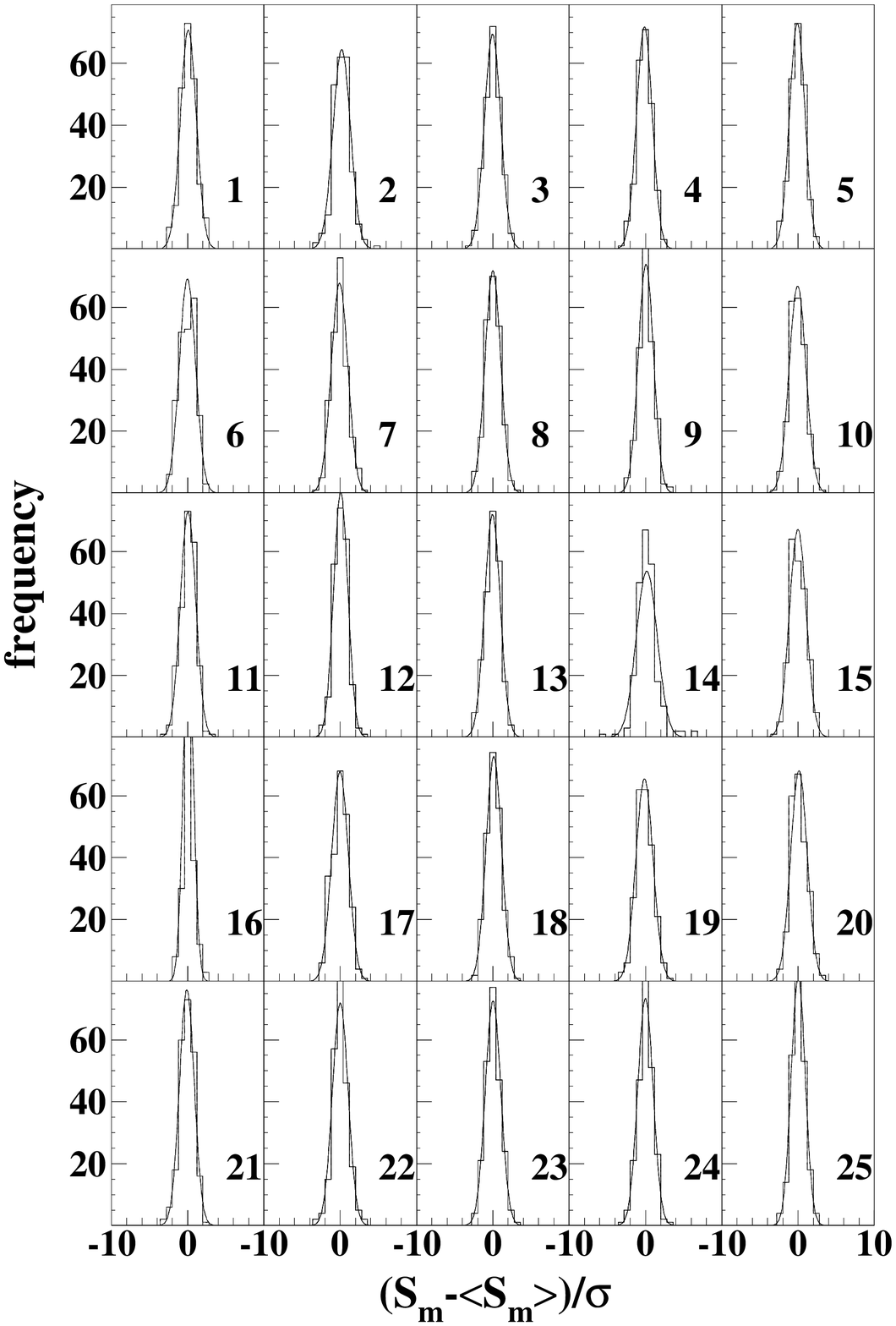}
\end{center}
\vspace{-0.8cm}
\caption{Histograms of the variable $\frac {S_m - \langle S_m \rangle}{\sigma}$, where
$\sigma$ are the errors
associated to the $S_m$ values and $\langle S_m \rangle$
are the mean values of the modulation amplitudes averaged over the detectors
and the annual cycles for each considered energy bin (here $\Delta E = 0.25$ keV).
Each panel refers to a single DAMA/LIBRA detector.
The entries of each histogram are 232
(the 16 energy bins in the (2--6) keV energy interval of the seven DAMA/LIBRA--phase1 annual cycles and 
 the 20 energy bins in the (1--6) keV energy interval of the six   DAMA/LIBRA--phase2 annual cycles),
but 152 for the 16$^{th}$ detector (see Ref. \cite{modlibra3}). 
The superimposed curves are gaussian fits.}
\label{fg:smcr}
\end{figure}

As previously done for the other data releases \cite{modlibra,modlibra2,modlibra3,review}, 
the $S_{m}$ values for each detector for each annual cycle and for each energy bin have been obtained. 
The $S_m$ are expected to follow a normal distribution in absence of any systematic effects.
Therefore, the variable $x = \frac {S_m - \langle S_m \rangle}{\sigma}$ has been considered
to verify that the $S_{m}$ are statistically well distributed 
in the 16 energy bins ($\Delta E = 0.25$ keV) in the (2--6) keV energy interval of the seven DAMA/LIBRA--phase1 annual cycles and
in the 20 energy bins in the (1--6) keV energy interval of the six DAMA/LIBRA--phase2 annual cycles and in each detector.
Here, $\sigma$ are the errors associated to $S_m$ and $\langle S_m \rangle$
are the mean values of the $S_m$ averaged over the detectors
and the annual cycles for each considered energy bin.
The distributions and their gaussian fits obtained for the detectors are shown in Fig.~\ref{fg:smcr}.

Defining $\chi^2 = \Sigma x^2$, where the sum is extended over
all the 232 (152 for the 16$^{th}$ detector \cite{modlibra3}),
$x$ values $\chi^2/d.o.f.$ values ranging from 0.69 to 1.95 are obtained.

The mean value of the 25 $\chi^2/d.o.f.$ is 1.07.
This value is slightly larger than 1.
Although this can be still ascribed to statistical fluctuations,
let us ascribe it to a possible systematics. In this case, one would
derive an additional error to the modulation amplitude
measured below 6 keV:
$\leq 2.1 \times 10^{-4}$ cpd/kg/keV, if combining quadratically the errors, or
$\leq 3.0 \times 10^{-5}$ cpd/kg/keV, if linearly combining them.
This possible additional error: $\leq 2\%$ or $\leq 0.3\%$, respectively, on the
DAMA/LIBRA--phase1 and DAMA/LIBRA--phase2 modulation amplitudes
is an upper limit of possible systematic effects coming from the detector to detector differences.

Among further additional tests, the analysis 
of the modulation amplitudes as a function of the energy separately for
the nine inner detectors and the remaining external ones has been carried out for DAMA/LIBRA--phase2,
as already done for the other data sets \cite{modlibra,modlibra2,modlibra3,review}. 
The obtained values are fully in agreement; in fact,
the hypothesis that the two sets of modulation amplitudes as a function of the
energy belong to same distribution has been verified by $\chi^2$ test, obtaining e.g.:
$\chi^2/d.o.f.$ = 2.5/6 and 
40.8/38 for the energy intervals (1--4) and (1--20) keV, 
respectively ($\Delta E$ = 0.5 keV). This shows that the
effect is also well shared between inner and outer detectors. 

\begin{figure}[!h]
\begin{center}
\vspace{-0.8cm}
\includegraphics[width=0.85\textwidth] {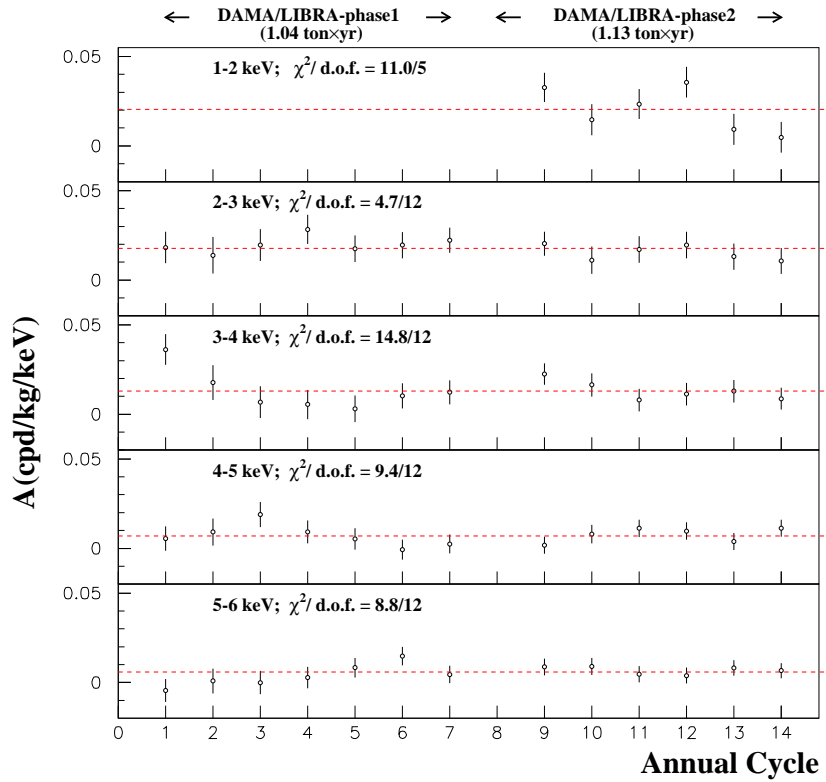}
\end{center}
\vspace{-0.7cm}
\caption{Modulation amplitudes of each single annual cycle
of DAMA/LIBRA--phase1 and DAMA/LIBRA--phase2. The error bars are the 1$\sigma$ errors.
The dashed horizontal lines show the central values obtained by best fit over
the whole data set.
The $\chi^2$ test and the {\it run test} accept the hypothesis at 95\% C.L. that the modulation amplitudes
are normally fluctuating around the best fit values.}
\vspace{-0.1cm}
\label{fg:ampall}
\end{figure}

In Fig.~\ref{fg:ampall} the modulation amplitudes singularly calculated for each 
annual cycle 
of DAMA/LIBRA--phase1 and DAMA/LIBRA--phase2 are shown. To test the hypothesis that the 
amplitudes are compatible and normally fluctuating around their mean values
the $\chi^2$ test has been performed.
The $\chi^2/d.o.f.$ values are also shown in Fig.~\ref{fg:ampall}; they corresponds to upper tail probability of 
5.2\%, 97\%, 25\%, 67\% and 72\%, respectively.
In addition to the $\chi^2$ test, another independent statistical test has been applied: the {\it run test} (see e.g. Ref. \cite{eadie});
it verifies the hypothesis that the positive (above the mean value) and negative (under the mean value) data points are randomly distributed.
The lower (upper) tail probabilities obtained by the {\it run test} are:
70(70)\%, 50(73)\%, 85(35)\%, 88(30)\% and 88(30)\%, respectively.
This analysis confirms that the data collected in all the annual cycles with 
DAMA/LIBRA--phase1 and phase2 are statistically compatible and can be considered together.

\section{Investigation of the annual modulation phase}

Let us, finally, release the assumption of the phase $t_0=152.5$ day in the procedure to 
evaluate the modulation amplitudes. In this case the signal can be alternatively written as:
\begin{eqnarray}
\label{eqn1} 
S_{i}(E) & = & S_{0}(E) + S_{m}(E) \cos\omega(t_i-t_0) + Z_{m}(E) \sin\omega(t_i-t_0) \\
        & = & S_{0}(E) + Y_{m}(E) \cos\omega(t_i-t^*).   \nonumber
\end{eqnarray}

\begin{figure}[!ht]
\begin{center}
\includegraphics[width=0.48\textwidth] {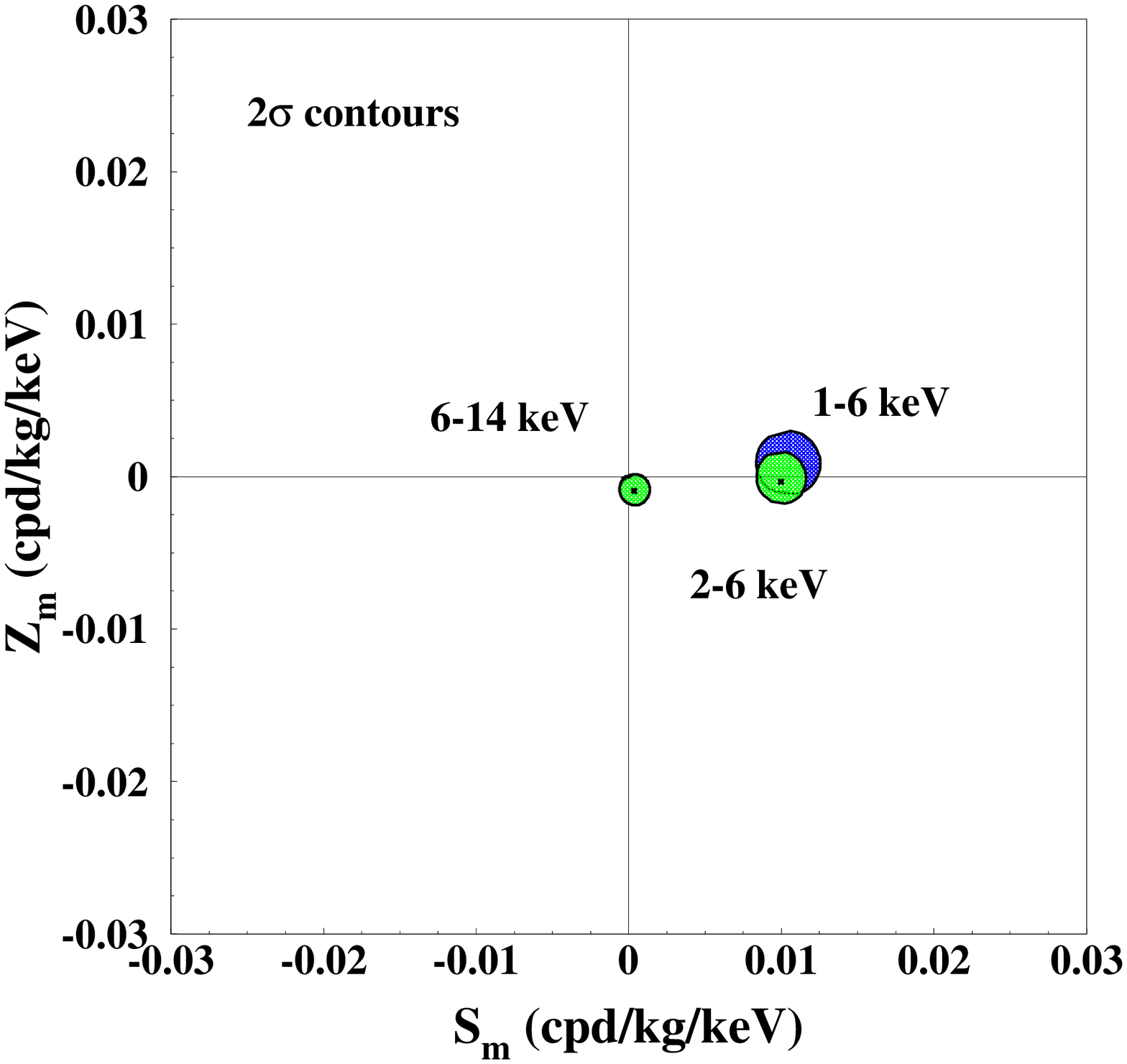}
\includegraphics[width=0.48\textwidth] {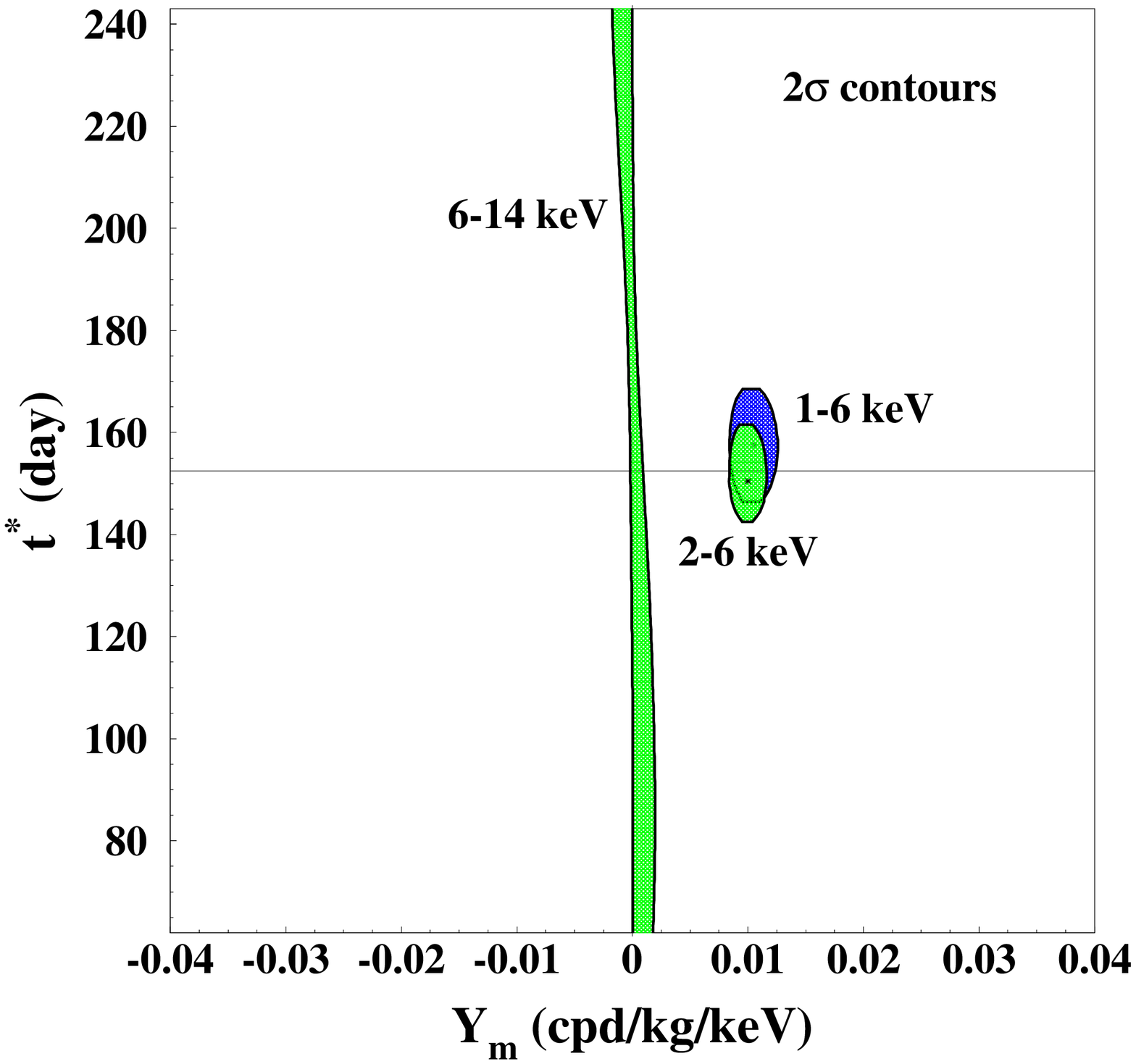}
\end{center}
\vspace{-0.7cm}
\caption{
$2\sigma$ contours in the plane $(S_m , Z_m)$ ({\it left})
and in the plane $(Y_m , t^*)$ ({\it right})
for:
i) DAMA/NaI, DAMA/LIBRA--phase1 and DAMA/LIBRA--phase2 in the (2--6) keV and (6--14) keV energy intervals (light areas, green on-line);
ii) only DAMA/LIBRA--phase2 in the (1--6) keV energy interval (dark areas, blue on-line).
The contours have been  
obtained by the maximum likelihood method.
A modulation amplitude is present in the lower energy intervals 
and the phase agrees with that expected for DM induced signals.}
\label{fg:bid}
\end{figure}

\noindent For signals induced by DM particles one should expect: 
i) $Z_{m} \sim 0$ (because of the orthogonality between the cosine and the sine functions); 
ii) $S_{m} \simeq Y_{m}$; iii) $t^* \simeq t_0=152.5$ day. 
In fact, these conditions hold for most of the dark halo models; however, as mentioned above,
slight differences can be expected in case of possible contributions
from non-thermalized DM components (see e.g. Refs. \cite{Fre04_1,Fre04_2,epj06,Fre05,Gel01,caus}).

\begin{table}[!ht]
\caption{Best fit values ($1\sigma$ errors) for  S$_m$ versus  Z$_m$ and  $Y_m$ versus $t^*$,
considering:
i) DAMA/NaI, DAMA/LIBRA--phase1 and DAMA/LIBRA--phase2 in the (2--6) keV and (6--14) keV energy intervals;
ii) only DAMA/LIBRA--phase2 in the (1--6) keV energy interval.
See also Fig.~\ref{fg:bid}.}
\begin{center}
\resizebox{\textwidth}{!}{
\begin{tabular}{|c||c|c||c|c|}
\hline
E     & S$_m$        & Z$_m$        & $Y_m$        & $t^*$  \\
(keV) & (cpd/kg/keV) & (cpd/kg/keV) & (cpd/kg/keV) & (day) \\
\hline
\multicolumn{5}{|l|}{DAMA/NaI+DAMA/LIBRA--phase1+DAMA/LIBRA--phase2:} \\
  2--6 &  (0.0100 $\pm$ 0.0008) & -(0.0003 $\pm$ 0.0008) &  (0.0100 $\pm$ 0.0008) & (150.5 $\pm$ 5.0) \\
 6--14 &  (0.0003 $\pm$ 0.0005) & -(0.0009 $\pm$ 0.0006) &  (0.0010 $\pm$ 0.0013) &  undefined       \\
\hline
\multicolumn{5}{|l|}{DAMA/LIBRA--phase2:} \\
  1--6 &  (0.0105 $\pm$ 0.0011) &  (0.0009 $\pm$ 0.0010) &  (0.0105 $\pm$ 0.0011) & (157.5 $\pm$ 5.0) \\
\hline
\hline
\end{tabular}}
\label{tb:bidbf}
\end{center}
\end{table}

\begin{figure}[!ht]
\begin{center}
\vspace{-1.2cm}
\includegraphics[width=0.95\textwidth] {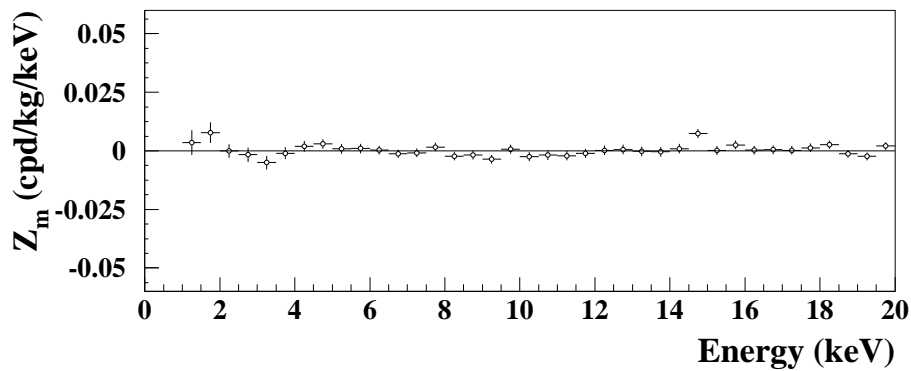}
\end{center}
\vspace{-0.8cm}
\caption{Energy distribution of the $Z_{m}$ variable for the cumulative exposure
of DAMA/NaI, DAMA/LIBRA--phase1, and DAMA/LIBRA--phase2 once setting $S_{m}$ in eq. (\ref{eqn1}) to zero.
The energy bin $\Delta E$ is 0.5 keV.
The $\chi^2$ test applied to the data supports the hypothesis that the $Z_{m}$ values are simply 
fluctuating around zero, as expected.}
\label{fg:zm}
\end{figure}

\begin{figure}[!ht]
\begin{center}
\includegraphics[width=0.85\textwidth] {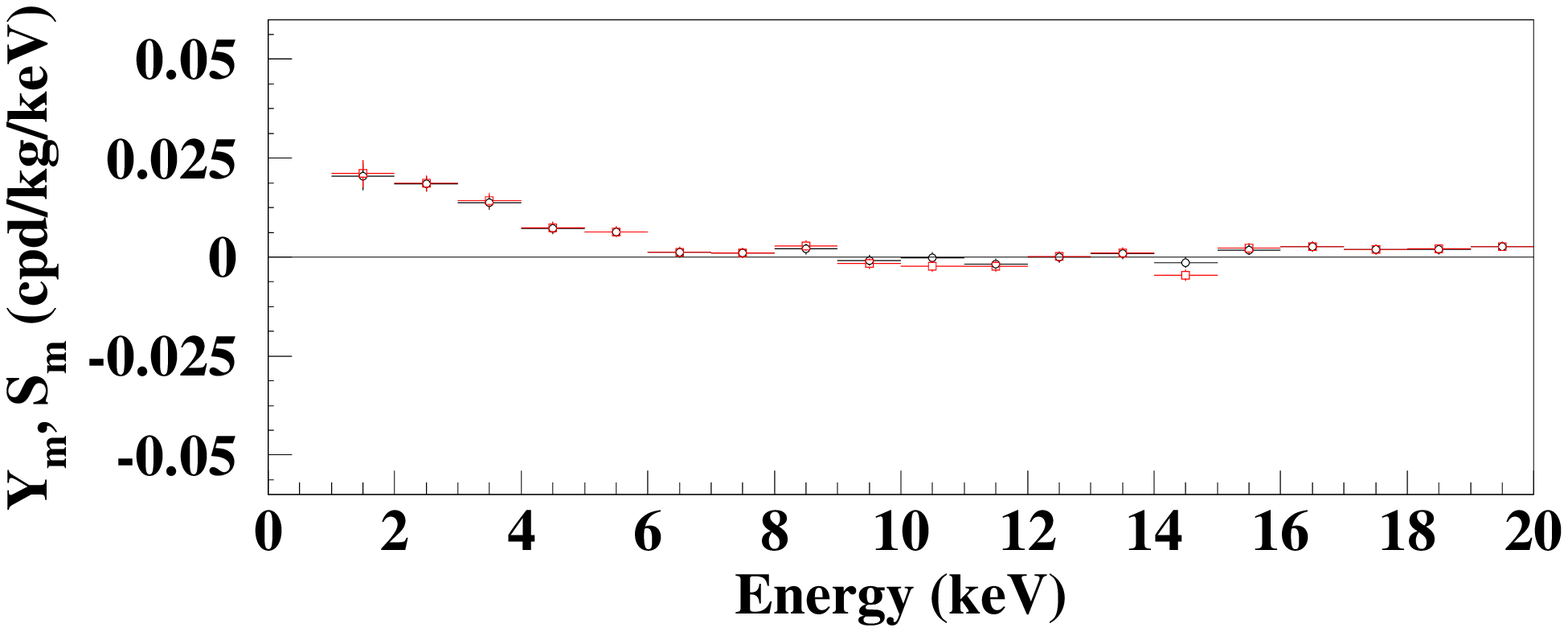}
\includegraphics[width=0.85\textwidth] {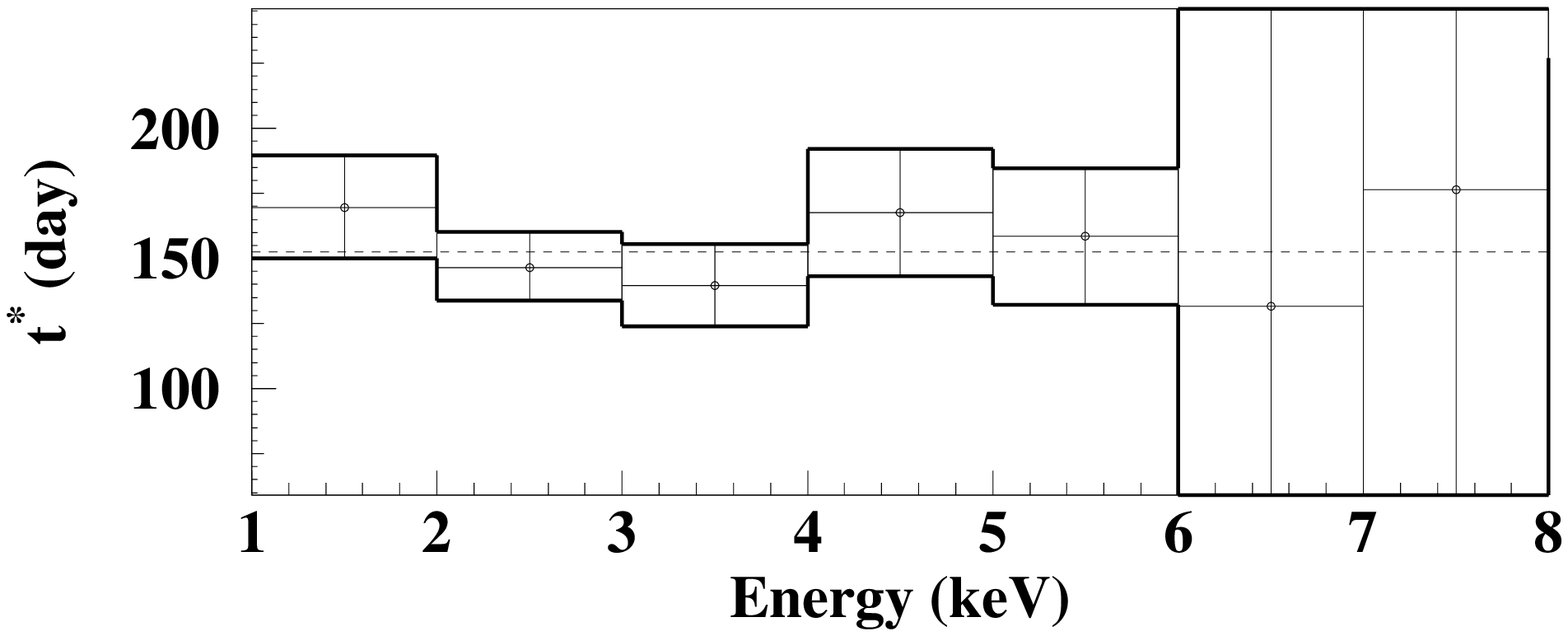}
\end{center}
\caption{{\em Top:} Energy distributions of the $Y_{m}$ variable (light data points; red color on-line)
and of the $S_{m}$ variable (solid data points; black on-line) for the cumulative exposure
of DAMA/NaI, DAMA/LIBRA--phase1, and DAMA/LIBRA--phase2. Here, unlike the data of
Fig.~\ref{fg:sme}, the energy bin is 1 keV.
{\em Bottom:} Energy distribution of the phase $t^*$ for the cumulative exposure
of DAMA/NaI, DAMA/LIBRA--phase1, and DAMA/LIBRA--phase2; here 
the errors are at $2\sigma$. The vertical scale spans over $\pm$ a quarter of period
around 2 June; other intervals are replica of it.
An annual modulation effect is present in the lower energy intervals 
up to 6 keV and the phase agrees with that expected for DM induced signals.
No modulation is present above 6 keV and thus the phase is undetermined.}
\label{fg:ymts}
\end{figure}

Considering cumulatively the data of DAMA/NaI, DAMA/LIBRA--phase1 and DAMA/LIBRA--phase2
the obtained $2\sigma$ contours in the plane $(S_m , Z_m)$ 
for the (2--6) keV and (6--14) keV energy intervals 
are shown in Fig.~\ref{fg:bid}{\it --left} while in 
Fig.~\ref{fg:bid}{\it --right} the obtained $2\sigma$ contours in the plane $(Y_m , t^*)$
are depicted.
Moreover, Fig.~\ref{fg:bid} also shows  only for DAMA/LIBRA--phase2 
the $2\sigma$ contours in the (1--6) keV energy interval.

The best fit values in the considered cases ($1\sigma$ errors) for S$_m$ versus Z$_m$ and $Y_m$ versus $t^*$ 
are reported in Table \ref{tb:bidbf}.

Finally, setting $S_{m}$ in eq. (\ref{eqn1}) to zero, 
the $Z_{m}$ values as function of the energy have also been determined
by using the same procedure. The $Z_{m}$ values
as a function of the energy are reported for
DAMA/NaI, DAMA/LIBRA--phase1, and DAMA/LIBRA--phase2 data sets
in Fig.~\ref{fg:zm}; they are expected to be zero. 
The $\chi^2$ test 
applied to the data supports the hypothesis that the $Z_{m}$ values are simply 
fluctuating around zero; in fact, 
in the (1--20) keV energy region the $\chi^2/d.o.f.$
is equal to 44.5/38 corresponding to a P-value = 22\%.

The energy behaviors of the $Y_{m}$ and of the phase $t^*$ 
are shown in Fig.~\ref{fg:ymts} 
for the cumulative exposure of DAMA/NaI, DAMA/LIBRA--phase1, and DAMA/LIBRA--phase2. 
The $Y_{m}$ are superimposed with the $S_{m}$ values 
with 1 keV energy bin (unlike Fig.~\ref{fg:sme} where the energy bin is 0.5 keV).
As in the previous analyses, an annual modulation effect is present in the lower energy intervals 
and the phase agrees with that expected for DM induced signals.
No modulation is present above 6 keV and the phase is undetermined.

\section{Conclusions}

The data of the new DAMA/LIBRA--phase2 confirm a
peculiar annual modulation of the {\it single-hit} scintillation events in the (1--6) keV energy region
satisfying all the many requirements of the DM annual modulation signature; the cumulative
exposure by the former DAMA/NaI, DAMA/LIBRA--phase1 and DAMA/LIBRA--phase2 is
2.46 ton $\times$ yr. 

As required by the exploited DM annual modulation signature: 
1) the {\it single-hit} events show a clear cosine-like modulation as expected for the DM signal; 
2) the measured period is equal to $(0.999\pm 0.001)$ yr well compatible with the 1 yr period as expected for the DM signal; 
3) the measured phase $(145\pm 5)$ days is compatible with the roughly $\simeq$ 152.5 days expected for the DM signal; 
4) the modulation is present only in the low energy (1--6) keV interval and not in other higher energy regions, consistently with expectation for the DM signal;
5) the modulation is present only in the {\it single-hit} events, while it is absent in the {\it multiple-hit} ones as expected for the DM signal;
6) the measured modulation amplitude in NaI(Tl) target of the {\it single-hit} scintillation events in the (2--6) keV energy interval, for which data are
also available by DAMA/NaI and DAMA/LIBRA--phase1,
is: $(0.0103 \pm 0.0008)$ cpd/kg/keV (12.9 $\sigma$ C.L.).
No systematic or side processes able to mimic the signature, i.e. able to 
simultaneously satisfy all the many peculiarities 
of the signature and to account for the whole measured modulation amplitude, has been 
found or suggested by anyone throughout some decades thus far.
In particular, arguments related to any possible role of some natural 
periodical phenomena have been discussed and quantitatively demonstrated 
to be unable to mimic the signature (see references; e.g. Refs. \cite{mu,norole}).
Thus, on the basis of the exploited signature, the model independent DAMA results give evidence 
at 12.9$\sigma$ C.L. (over 20 independent annual cycles and in various experimental configurations) for 
the presence of DM particles in the galactic halo.

\vspace{0.2cm}

In order to perform corollary investigation on the nature of the DM particles in given scenarios, model-dependent 
analyses are necessary\footnote{It is worth noting that it does not exist in direct and indirect DM detection experiments 
approaches which can offer such information independently on assumed models.}; thus,
many theoretical and experimental parameters and models are possible
and many hypotheses must also be exploited. In particular, the DAMA model independent evidence is compatible with a wide 
set of astrophysical, nuclear and particle physics scenarios 
for high and low mass candidates inducing nuclear recoil and/or electromagnetic radiation,
as also shown in various literature.
Moreover, both the negative results and all the possible positive hints, achieved so-far
in the field, can be compatible with the DAMA model independent DM annual modulation results in many scenarios considering also the 
existing experimental and theoretical uncertainties; the same holds for indirect approaches. 
For a discussion see e.g. Ref.~\cite{review} and references therein. 
Updated/new corollary analyses on various possible DM 
scenarios will be addressed in the next dedicated works.

Finally, we stress that to efficiently disentangle among at least some of the many possible candidates and scenarios 
an increase of exposure in the new lowest energy bin is important. The experiment is collecting data 
and related R\&D is under way.

\section{Acknowledgments} 

This paper is dedicated to the memory of Prof. L. Paoluzi, Director of the INFN-Roma2 at time of starting the DAMA 
project, of Prof. D. Prosperi, one of the main proponents of this project, and of Prof. S. d'Angelo who worked in 
some DAMA activities, and always gave us fruitful scientific and personal support.

The authors also gratefully acknowledge the presidents of the Scientific Committee II of the I.N.F.N. and the 
referees of the DAMA project there, along various periods. They also wish to thank all the technical staffs who 
supported the works, and all the colleagues who contribute
to the various searches on rare processes with the DAMA low-background set-ups.

\end{document}